\documentclass[14pt]{article}

\usepackage{amsmath,graphicx,amsfonts}
\usepackage{epsfig}
\usepackage{subfigure}
\usepackage{dcolumn}

\usepackage{amsfonts}
\usepackage{amssymb}
\usepackage{graphicx}
\def\siz{\small}
\def\be{\begin{equation}}
\def\ee{\end{equation}}

\usepackage{soul}
\usepackage{amssymb,amsmath}
\usepackage{graphicx}
\usepackage{hyphenat}
\begin{document}

\title{Nonstandard approach to gravity for the dark sector of the Universe\footnote{Accepted for publication by `Entropy' in the special issue `Modified gravity: From black holes entropy to current cosmology'}}
\author{P.C. Stichel$^{1)}$  and W.J.
Zakrzewski$^{2)}$
\\
\siz
$^{1)}$An der Krebskuhle 21, D-33619 Bielefeld, Germany \\ \siz
e-mail:peter@physik.uni-bielefeld.de
\\ \\ \siz
$^{2)}$Department of Mathematical Sciences, University of Durham, \\
\siz Durham DH1 3LE, UK \\ \siz
 e-mail: W.J.Zakrzewski@durham.ac.uk
 }

\date{}
\maketitle

\abstract{
We summarize the present state of research on the darkon fluid as a model for the dark sector of the Universe. Nonrelativistic massless particles are introduced as a realization of the Galilei group in an enlarged phase space. The additional degrees of freedom allow for a nonstandard, minimal coupling to gravity respecting Einstein's equivalence principle. Extended to a self-gravitating fluid the Poisson equation for the gravitational potential contains a dynamically generated effective gravitational mass density of either sign. The equations of motion (EOMs) contain no free parameters and are invariant w.r.t.  
Milne gauge transformations.  Fixing the gauge eliminates the unphysical degrees of freedom. The resulting Lagrangian possesses no free particle limit. The particles it describes, darkons, exist only as fluid particles of a self-gravitating fluid.
 This darkon fluid realizes the zero-mass Galilean algebra extended by dilations with dynamical exponent $z = \frac{5}{3}$. We reduce the EOMs to Friedmann-like equations, derive conserved quantities and a unique Hamiltonian dynamics by implementing dilation symmetry. By the Casimir of the Poisson-bracket (PB)-algebra we foliate the phase space and construct a Lagrangian in reduced phase space. We solve the Friedmann-like equations with the transition redshift and the value of the Casimir as integration constants. We obtain a deceleration phase for the early Universe and an acceleration phase for the late Universe in agreement with observations. Steady state equations in the spherically symmetric case may model a galactic halo. Numerical solutions of a nonlinear differential equation for the gravitational potential lead to predictions for the dark matter (DM) part of the rotation curves (RCs) of galaxies in qualitative agreement with observational data. We also present a general covariant generalization of the model.} 





\eject
\centerline{\Large Contents}
\begin{itemize}                                          
\item 1 Introduction   \hfill 3                                                             
\item 2 Galilean massless particles and their nonstandard coupling to gravity\hfill 5
\begin{itemize}
\item    2.1 Galilean massless particles \hfill 5
\item     2.2 Coupling to gravity  \hfill 8
\end{itemize}
\item  3 A self-gravitating (darkon) fluid  \hfill 9
\begin{itemize}
\item     3.1 Lagrangian formulation    \hfill 9
\item     3.2 Relabeling symmetry and the transport equation for the gravitational field \hfill 11
\item      3.3 Eulerian formulation  \hfill 11
\item     3.4 Coupling with baryonic matter   \hfill 12
\end{itemize}
\item  4  Hamiltonian dynamics   \hfill 13
\item  5  Space-time symmetries   \hfill 15
\begin{itemize}
\item      5.1 Space-time translations and the energy-momentum tensor  \hfill 16
\item      5.2 Zero mass Galilean symmetry  \hfill 16
\item      5.3 Angular momentum    \hfill 17
\item     5.4 Anisotropic scaling    \hfill 17
\end{itemize}
\item  6   Darkon fluid cosmology    \hfill 18
\begin{itemize}
\item       6.1 Cosmological EOMs    \hfill 18
\item       6.2 Hamiltonian dynamics   \hfill 19
\item      6.3 Solution of the EOMs   \hfill 24
\item      6.4 Distances    \hfill 26
 \item     6.5 Predictions versus observations   \hfill 27
\item      6.6 Comparison with alternative models for dark energy   \hfill 29
\end{itemize}
\item 7  Modeling dark matter halos by a steady state darkon fluid   \hfill 31
\begin{itemize}
\item      7.1 Equation for the gravitational potential   \hfill 31
\item      7.2 Predictions versus observations    \hfill 34
\end{itemize}
\item 8  Influence  of the cosmic expansion on binary systems   \hfill 38
\item 9  Outlook for a general covariant theory   \hfill 40
\item 10  Conclusions and outlook \hfill 41
\end{itemize} 
\eject

\section{Introduction}

Understanding the dark sector of the Universe is one of the greatest challenges of today's theoretical physics. The main question is whether the underlying astrophysical observations (missing gravitation in galactic systems resp. the accelerated expansion of the Universe) can be understood by means of known physical concepts, {\it e.g.} General Relativity (GR), or whether some kind of ``new'' physics is needed. Indeed, cosmic acceleration could be an apparent effect due to the averaging over large scale inhomogeneities in the Universe, but a reliable quantification of such an effect is not yet available (see the recent review \cite{a1}). On the other hand there exists an overwhelming evidence for the existence of gravitational effects on all cosmological scales (termed ``dark matter'' (DM)) which cannot be explained by the gravitation of standard matter in the framework of GR (see the review \cite{a2}).

 The current picture of cosmological structure formation assumes for DM some pressure less dust containing massive, perhaps weakly interacting cold particles (CDM). The present standard cosmological model ($\Lambda$CDM-model) uses besides CDM a positive cosmological constant  $\Lambda$ as the cause for the accelerated expansion of the Universe (see recent reviews on dark energy (DE), {\it e.g.} \cite{a3}, \cite{a4} and \cite{a5}).  But, ``favoured by a number of observations'' \cite{a3}, the $\Lambda$CDM-model suffers at least from the following insufficiencies:
\begin{itemize}
\item Interpreted as the energy density of the vacuum, the experimental value of $\Lambda$ turns out to be a factor of about $10^{54}$ too small (see the recent estimate in \cite{a6}).
\item None of the proposed CDM-constituents has been observed (cp. \cite{a7}).
\item Rotation curves (RCs) of DM-dominated galaxies behave in their inner part in sharp contrast to the CDM-based simulations (known as the core-cusp problem \cite{a8}).
\end{itemize}

In a very recent paper P. Kroupa \cite{a9} lists a large set of extragalactic observations which falsify the $\Lambda$CDM-model.

In this paper we will neither comment on the huge number of phenomenological models for a dynamical DE (see \cite{a3}-\cite{a5}), which all rest on at least one unknown function ({\it e.g.} a scalar field potential), nor on those modified gravity theories which explain DM-effects only ({\it e.g.} modified Newtonian dynamics (MOND) in its simplest, Milgromian form \cite{a10}). But we know of at least two kinds of modified gravity models which explain both, DE and DM {\it i.e.} the whole dark sector of the Universe:
\begin{itemize}
\item For MOND-models see section 9 of the very recent review \cite{a11}. But all these models are very phenomenological in that they are based on one unknown ``MOND function''.
\item Modified theory of gravity (MOG) by Moffat \cite{a12} in which, besides the Einstein-Hilbert and matter actions, a massive vector field is introduced whose mass, coupling constant to matter and gravitational constant are promoted to scalar fields. The self-interaction potentials of these four fields are not fixed a priori and play the role of DE \cite{a13}. This model can explain galactic phenomena as {\it e.g.} RCs \cite{a14} as well as the accelerated expansion of the Universe and other cosmological observations \cite{a13}.
\end{itemize}

In summary, all present attempts to explain quantitatively the new cosmological observations ({\it e.g.} late time cosmic acceleration) as well as the galactic RCs and other galactic phenomena contain either some new parameters or even free functions. If we assume that the  cosmic acceleration is a real effect and not an apparent one (see above) then, obviously, we need some new ({\it i.e.} unconventional) physics which, however, should be based on known fundamental physical principles ({\it e.g.} symmetry).

 In this paper we review some ongoing research on a model which is a first building block for  a new theory describing the dark sector of the Universe. This model, introduced in \cite{a15} and further developed in \cite{a16} and \cite{a17}, contains no new parameters in its Lagrangian and is based on well- known physical principles (Galilei symmetry and Einstein's equivalence principle). The only free parameters appearing in our model are some integration constants (see sections 6 and 7). The basic idea it uses involves nonrelativistic massless particles. The use of such particles  may appear very strange at a first sight as  it seems to contradict Special Relativity according to which massless particles must move with the velocity of light. However, as we have shown before, we can introduce nonrelativistic massless particles if we enlarge the dimension of phase space; {\it i.e.} such particles are not described by position and momentum only.  The additional degrees of freedom allow for a nonstandard, minimal coupling of these ``exotic'' particles \cite{a15} to gravity in accordance with Einstein's equivalence principle. This coupling, for a self-gravitating fluid, leads to a dynamically generated effective gravitational mass density of either sign which is the source of the corresponding gravitational field. This fact leads to the possibility of using such a model as a first building block of a new theory describing the dark sector.  In our approach we formulate the equivalence principle (local equivalence between gravitation and acceleration) as usual as an invariance of the dynamics w.r.t. arbitrary time-dependent translations (Milne gauge transformations). By fixing the gauge we can reduce the dimension of phase space and remove the unphysical variables. The Lagrangian of the resultant model does not possess a free particle limit and hence the particles it describes, called darkons, exist only as fluid particles of a self-gravitating fluid. This darkon fluid presents a dynamical realization of the zero-mass Galilean algebra extended by anisotropic dilations with the dynamical exponent $z = \frac{5}{3}$. This dilation symmetry emerges in a natural way from the minimal gravitational coupling.
The cosmological EOMs are derived, as usual, from the darkon fluid EOMs by restricting the form of their solutions through the cosmological principle.  However,  the corresponding Hamiltonian cannot be derived from the darkon fluid Hamiltonian. So we have to derive a Hamiltonian and the corresponding Poisson-brackets (PBs) from the EOMs, a procedure which, as is well known, is not unique.  So we construct, using the constants of motion, a Hamiltonian which has the correct scaling dimension and obtain the unique PBs by implementing the dilation symmetry. Moreover, our procedure allows us to construct a Lagrangian having foliated the phase space by means of the Casimir of the PB-aLgebra. We solve the cosmological EOMs and obtain, by fixing two integration constants by their values determined from experimental data, a prediction for the Hubble parameter $H(z)$ ($z$ is the redshift) in general agreement with the existing data.

Furthermore, we demonstrate that our model not only gives a proper description of the late time cosmic acceleration but it also predicts galactic halos and, qualitatively correct, the corresponding flat rotation curves (RCs).  To do this we derive for the spherically symmetric steady state case a nonlinear ordinary differential equation for the gravitational potential which is then solved numerically. 

Almost all cosmological models are based on General Relativity (GR) or on some of its modifications which, however, preserve general covariance. The appearance of anisotropic dilations in our model seems, at first sight, to forbid a generalization of our model to a general covariant theory. But the dilation symmetry with $z=\frac{5}{3}$ is an emergent symmetry and not an input of our model. So we are able to propose a generalization of our model showing general covariance without imposing any form of scale symmetry.

One of the aims of the present paper is to show in detail the main ideas of our model. A comparison with observational data (Hubble parameter resp. galactic rotation curves) serves only as an illustration that it might be meaningful to develop the present model to a proper theory of gravitation for the dark sector of the Universe. For this reason we do not produce any least square fits for the undetermined integration constants appearing in the cosmological context.

The paper is organized as follows: In section 2 we introduce Galilean massless particles and their coupling to gravity. In section 3 we describe the generalization of this particle picture to a self-gravitating fluid, fix the gauge and so obtain the darkon fluid. Then, instead of the Poisson equation, we get a transport equation for the gravitational field. In section 4 we describe the Hamiltonian dynamics of the darkon fluid  to have a basis for quantization.  Space-time symmetries of our model are discussed in section 5. The darkon fluid cosmology is presented in section 6. Starting with the cosmological principle we derive Friedmann-like equations, discuss the corresponding Hamiltonian dynamics and compare our predictions with the existing data for the Hubble diagram. In section 7 we consider steady state equations in the spherically symmetric case and derive a nonlinear, ordinary differential equation (ODE) for the gravitational potential. Numerical solutions of this ODE lead to predictions for the behaviour of the DM-part of galactic RCs. In section 8 we outline how to determine a correction to Newton's gravitational law due to the cosmic expansion. Finally, in section 9 we present a general covariant generalisation of our model.
We close with some conclusions and give an outlook for future work.  

\section{Galilean massless particles and their nonstandard coupling to gravity.}

As stated earlier the introduction of nonrelativistic massless particles may appear very strange at first sight. It seems to contradict Special Relativity according to which massless particles must move with the velocity of light.  But in Special Relativity spinless particles are described by the vectors of position and momentum only. In our recent papers \cite{a15}-\cite{a17}  we have shown that we can introduce nonrelativistic massless particles if we enlarge the dimension of the phase space from six to twelve.  In subsection 2.1 we will show that this enlargement of phase space is also necessary: The 12-dimensional phase space is the minimal one allowing the dynamical realization of nonrelativistic massless particles.

How to couple gravity if there is no mass? Our nonrelativistic framework forbids the coupling to  energy, as then this would be a relativistic effect. 
 So we take  Einstein${'}$s equivalence principle as a starting point in subsection 2.2. Then, within the Lagrangian framework, the gravitational field will be coupled to one of the new degrees of freedom. 
\subsection{ Galilean massless particles}

Our first task involves finding a minimal dynamical realization of the unextended ($m=0$) Galilean algebra  $G_0$         in three space dimensions.

 $G_0$   is given by the Lie-brackets

$$ [A_i,\,L_j]\,=\,\epsilon_{ijk}\,A_k,\quad A_i\in(P_i,K_i,L_i),$$
\be                                                                                                                                                                              
[H,K_i]\,=\,P_i,\quad [H,\,P_i]\,=\,[H,\,L_i]\,=\,0,
\label{2a}
\ee
$$[P_i,\,K_j]\,=\,0\quad [P_i,\,P_j]\,=\,0,\quad [K_i,\,K_j]\,=\,0, $$
where  $H,P_i,K_i$   and  $L_i$  are the generators of time-resp. space translations, boosts and rotations.

A minimal dynamical realization of (\ref{2a}) involves a  construction of the elements of  $G_0$  in terms of a minimal number of phase space variables and without the introduction of any free functions or parameters. Then the Lie-brackets in (\ref{2a}) have to be understood as Poisson-brackets (PBs) w.r.t. the PB-algebra of phase space variables.

Let $\vec x(t)$   be a particle trajectory and  $\vec p$  the corresponding conjugate momentum satisfying the fundamental PB-relation
\be
[x_i,\,p_j]\,=\,\delta_{ij}
\label{2b}
\ee                                                                                                                                                                                 
The translation generator  $\vec P$  acts on  $\vec x$  resp.  $\vec p$     as
$$ [x_i,\,P_j]\,=\,\delta_{ij}\quad \hbox{resp}\quad [p_i,\,P_j]=0.
$$

Therefore we get the identification
 \be
P_i\,=\,p_i.
\label{2c}
\ee 
                                                                                                                                                                              
To show that our minimally enlarged phase space contains, besides $\vec x$ and $\vec p$,
two additional vectors corresponding to the reduced boost $\vec q$ and the velocity $\vec y$, we proceed in two steps:
\begin{itemize}

\item 1. Define a reduced boost vector $\vec q$
\be
q_i\equiv K_i\,-\,p_it.
\label{2d}
\ee                                                                                                                                                                                
        The boost $\vec K$       acts on $\vec x$         as
  \be [x_i,\,K_j]\,=\,\delta_{ij}t
\label{2e}
\ee                                                                                                                                                                              
        leading by (\ref{2b}) and (\ref{2d}) to
 \be
[x_i,\,q_j]\,=\,0.
\label{2f}                                                                                                                                                                               
\ee
  Taking  the PB of (\ref{2d}) with    $p_j$        it is easily seen by the expressions in (\ref{2a})  that the PB of $p_i$ and $q_j$ vanishes 
 \be [p_i,\,q_j]\,=\,0.
\label{2g}
\ee  
      
Despite of (\ref{2f}) and (\ref{2g})  $\vec q$            turns out to be nontrivial: Taking the PB of (\ref{2d}) with $H$ gives
\be
\dot q_i\,=\,[q_i,\,H]\,=\,-p_i.
\label{2h}\ee

Thus we conclude from (\ref{2f})-(\ref{2h}) that  the $q_i$   are independent dynamical variables which, according to (\ref{2d}) and (\ref{2a}), are commutative
 \be
[q_i,\,q_j]\,=\,0.
\label{2ha}
\ee
 
\item 2. Next we introduce the velocity vector
 \be
y_i\equiv [x_i,\,H],\label{2i}
\ee                                                                                                                                                       
which is translationally invariant (take the PB of (\ref{2i}) with  $p_j$         and use the Jacobi identity) to get
\be
[y_i,\,p_j]\,=\,0.
\label{2j}
\ee

From (\ref{2i}), (\ref{2a}), (\ref{2b}) and (\ref{2e}) we get the action of the boost generator on $y_i$    
\be
[y_i,\,K_j]\,=\,\delta_{ij}.
\label{2k}
\ee

Inserting now (\ref{2d}) into (\ref{2k}) leads by (\ref{2j}) to                        
\be
[y_i,\,q_j]\,=\,\delta_{ij}.\label{2l}                              
\ee
We conclude from (\ref{2l}) and by (\ref{2f}), (\ref{2g}) and (\ref{2ha})                           
 that  $\vec y$      cannot  be a function
of the other variables $\vec x$, $\vec p$ or $\vec q$.)

To close the PB-algebra in phase space we choose, in accordance with the principle of a minimal dimension, 
\be 
[x_i,\,y_j]\,=\,0.
\label{2n}
\ee
Note that (\ref{2n}) respects all Jacobi identities.
\end{itemize}

It remains to construct out of the phase space variables $\vec x$, $\vec p$, $\vec q$, and $\vec y$ the Hamiltonian 
$H$ and the generator of rotations $\vec L$.

 Our foregoing results lead to the EOMs
\be
\dot x_i\,=\,y_i,\quad \dot p_i\,=\,0\quad \hbox{and}\quad \dot q_i\,=\,-p_i
\label{2p}
\ee
but $\dot y_i$  has not not yet been fixed.

The EOMs (\ref{2p}) together with the canonical PBs 
\be
[x_i,\,p_j]\,=\,\delta_{ij},\quad [y_i,\,q_j]\,=\,\delta_{ij}
\label{2q}\ee
imply that the Hamiltonian is given by
$$
H\,=\,p_iy_i\,+\,f(\vec q)$$
The principle of a minimal dynamical realization of $G_0$      forces us to put $f=0$. Thus we see that the Hamiltonian 
$H$ is given by
\be
H\,=\,p_iy_i,
\label{2o}
\ee
which leads, due to (\ref{2q}), to the  EOM for $y_i$, namely
\be
\dot y_i\,=\,0.
\label{2ma}
\ee
The conserved generator of rotations $L_i$ is then given by
\be
L_i\,=\,\epsilon_{ikl}\,(x_kp_l\,+\,y_kq_l).
\label{2na}
\ee
The two parts of $L_i$ act separately on the $(\vec x,\vec p)$, resp. $(\vec y,\vec q)$ parts of the phase space.

The Hamiltonian dynamics given by $H$ (\ref{2o}) and the nonvanishing PBs (\ref{2q}) is equivalent to the dynamics derived from the Lagrangian
\be 
L_0\,=\,p_i\dot x_i\,+\,q_i\dot y_i\,-\,p_iy_i.
\label{2pa}
\ee

Finally we conclude
\begin{itemize}
\item A minimal dynamical realization of the unextended Galilei algebra in three dimensional configuration space requires a 12-dimensional phase space  spanned by $(\vec x,\,\vec p,\,\,\vec q,\,\vec y)$
\item The relation between energy $H$ and the momenta (velocities) is nonstandard,
\item Momenta    $p_i$     and velocities  $y_i$       are dynamically independent,
\item The $q_i$       have no counterpart in standard particle mechanics,
\item Galilean massless particles move with arbitrary finite velocity.
\end{itemize}

{\bf Dilations}

  $G_0$     may be enlarged by dilations with arbitrary dynamical exponent  $z$. The corresponding generator has to satisfy the following PB-relations (cp \cite{a} for $z=\frac{N}{2}$)
 \be
[D,\,H]\,=\,-H,\quad [D,\,P_i]\,=\,-\frac{1}{z}P_i,
\label{2qa}
\ee
$$[D,\,K_i]\,=\,\left(1-\frac{1}{z}\right)K_i,\quad [D,\,L_i]\,=\,0.
$$

It can easily be checked that    $D$         is given by 
\be
D\,=\,tH\,-\,\frac{1}{z}x_ip_i\,+\,\left(1-\frac{1}{z}\right)\,y_iq_i.
\label{2r}
\ee

Below we will show that the value of  $z$ becomes fixed by the coupling of our massless particles to gravity.

\subsection{Coupling to gravity}
 
To couple our massless particles to the gravitational field strength  $g_i(\vec x,t)$            we have to be consistent with Einstein$'$s equivalence principle:  Locally, the gravitational field strength is equivalent to an accelerating frame. The only known nonrelativistic EOM for the particle trajectory       satisfying this form of the equivalence principle is given by the Newton law:
\be 
\ddot x_i(t)\,=\,g_i(\vec x(t),t)
\label{2s}
\ee
because (\ref{2s}) is invariant w.r.t. arbitrary time-dependent translations 
(Milne gauge transformations) \cite{b}                                                                      
\be
x_i\,\quad \rightarrow \quad x_i^{'}\,=\,x_i\,+\,a_i(t)
\label{2t}
\ee
provided that $g_i$           transforms to
 \be g^{'}_i(\vec x^{'},t)\,=\,g_i(\vec x,t)\,+\,\ddot a_i(t).
\label{2u}
\ee                                                                                                                                                                             
The EOM (\ref{2s}) is realized if we add to $L_0$  an interaction part (minimal coupling)
\be
L_{int}\,=\,-q_i\,g_i(\vec x,t)
\label{2w}\ee

Then the EOM   $\ddot q_i=0$      gets replaced by
\be
\ddot q_i\,=\,q_k\partial_i g_k.
\label{2y}\ee

The total Lagrangian $L=L_0+L_{int}$  now becomes invariant  w.r.t. the gauge transformations   
\be
x_i^{'}\,=\,x_i\,+\,a_i(t),\quad,y_i^{'}\,=\,y_i\,+\,\dot a_i(t),\quad p^{'}_i\,=\,p_i,\quad q_i^{'}\,=\,q_i,
\label{2z}
\ee
$$g^{'}_i(\vec x^{'},t)\,=\,g_i(\vec x,t)\,+\,\ddot a_i(t).$$

This gauge invariance, resp. Einstein$'$s equivalence principle still holds if we add in (\ref{2w}) further terms containing spatial derivatives of  $g_i$         {\it i.e.} by replacing
 \be                                                                        
g_i\,\rightarrow\,g_i\,-\,\kappa_1\triangle g_i\quad                                                                                                                   +\hbox{higher}\quad \hbox{order}\quad \hbox{terms},
\label{2za}
\ee
where $\triangle$ is the laplacian and $K_1$ is a constant.

\section{  A self-gravitating (darkon) fluid}
\subsection{Lagrangian formulation}

In the last subsection (2.2) we have introduced a model for the motion of one Galilean massless particle in an external gravitational field $\vec g(\vec x,t)$.  
              
Now we want to generalize this model in two respects:
\begin{itemize}
\item  1. Instead of one particle we consider a continuum (fluid) labeled by comoving coordinates $\vec \xi\in \mathbb{R}^3$                (Lagrangian formulation of fluid dynamics, cp. \cite{c}).

\item 2. We treat the gravitational field  $\vec g$      as a  dynamical variable in order to get a self-gravitating fluid.
\end{itemize}

 To realize the first point we substitute in the EOMs (\ref{2s}) and (\ref{2y})
\be
x_i(t)\quad \rightarrow \quad x_i(\vec \xi,t)\quad \hbox{and}\quad q_i(t)\quad \rightarrow\quad q_i(\vec \xi,t).
\label{3a}\ee
 
To realize the second point we add to our previous  Lagrangian  $L_0+L_{int}$     a kinetic part  $L_{field}$                    for the gravitational field. We choose as usual
\be
L_{field}\,=\,-\frac{1}{8\pi G}\,\int \,d^3x\,(g_i(\vec x,t))^2,
\label{3b}
\ee
where  $G$ is Newton's gravitational constant.

Then our Lagrangian, formulated in terms of  $(x_i,\dot x_i),(q_i,\dot q_i)$                           and   $\vec g$,
becomes
\be
L\,=\,-n_0\,\int\,d^3\xi\,\left(\dot q_i(\vec \xi,t)\dot x_i(\vec \xi,t)\,+\,q_i(\vec \xi,t)g_i(\vec x(\vec \xi,t),t)\right)\,+\,L_{field},
\label{3c}\ee
where $n_0$        is the (constant) particle density in  $\vec \xi$-space.

Varying    $q_i$ ,         resp.,   $x_i$          leads to the continuum generalization of the previous EOMs
(\ref{2s}) and (\ref{2y})
\be
\ddot x_i(\vec \xi,t)\,=\,g_i(\vec x(\vec \xi,t),t)
\label{3d}
\ee
and
\be
\ddot q_i(\vec \xi,t)\,=\,q_k(\vec \xi,t)\,\frac{\partial}{\partial x_i}\,g_k(\vec x(\vec \xi,t),t).
\label{3e}\ee

Now we have {\bf two possibilities  to go further}:

In the {\bf first case} we take $g_i=-\partial_i \phi$ and vary $\phi$ getting
\be
\triangle \phi\,=\,4\pi G\,\partial_k(nq_k),
\label{3f}
\ee
with
\be
n(\vec x,t)\,=\,n_0\,J^{-1}\vert_{\vec\xi=\vec\xi(\vec x,t)}\quad \hbox{(particle}\quad\hbox{density)}
\label{3g}
\ee
and
\be
J\,=\,\det\left(\frac{\partial x_i(\vec \xi,t)}{\partial \xi_k}\right)\quad \hbox{(Jacobian)},
\label{3h}
\ee
where
$\vec \xi(\vec x,t)$ denotes the function inverse to $\vec x(\vec \xi,t)$.

The right hand side of the Poisson equation (\ref{3f}) describes a dynamically generated effective gravitational mass density which may be of either sign with

+ sign leading to an attractive gravitation,

- sign leading to a repulsive gravitation. 

This promotes the self-gravitating fluid to being a possible candidate for the dark sector of the Universe.

Note that our EOMs are invariant w.r.t the gauge transformations (see (\ref{2t}), (\ref{2u}))
\be
x_i\quad \rightarrow \quad x_i\,+\,a_i(t),
\label{3i}
\ee
when $\phi$ transforms as
\be
\phi(\vec x,t)\quad \rightarrow\quad \phi^{'}(\vec x^{'},t)\,=\,\phi(\vec x,t)\,-\,\ddot a_i(t)x_i,
\label{3j}
\ee
and $q_i$ and $n$ remain invariant.

In the {\bf second case} we vary $g_i$ instead of $\phi$ and this leads to a linear relation between $q_i$ and $g_i$:
\be
g_i(\vec x,t)\,=\,-4\pi \,G\,n(\vec x,t)\,q_i(\vec \xi(\vec x,t),t).
\label{3k}
\ee

Note that the expression for $g_i$ can be rewritten as
\be
q_i(\vec \xi,t)\,=\,-\frac{1}{4\pi G}\left(\frac{g_i}{n}\right)(\vec x(\vec \xi,t),t).
\label{3l}
\ee

The expression (\ref{3l}) allows to eliminate, from the EOMs, the unphysical phase space variables $q_i$ in favour of the gravitational field strength $g_i$.

This removes the unphysical degrees of freedom. However, we have to pay a prize for this 
 as now
\begin{itemize}
\item the gravitational field strength is not automatically given by the gradient of a potential,
\item  the expression (41) destroys the gauge symmetry {\it i.e.} it fixes the gauge.
\end{itemize}

{\bf Remark}  The gauge symmetry can be restored by considering a curl-free velocity field
$$
\vec u\,=\,\vec \nabla u$$
and treating $u$ as the variable to be varied \cite{a16}.

Inserting $q_i$ from (\ref{3l}) into the previous EOMs we get two coupled EOMs for $\vec x(\vec \xi,t)$ and $\vec g(\vec x(\vec \xi,t),t)$ 
\be
\frac{d^2}{dt^2}\,x_i(\vec \xi,t)\,=\,g_i(\vec x(\vec \xi,t),t)
\label{3m}
\ee
(initial condition $\vec x(\vec \xi,0)=\vec \xi$) and
\be
\frac{d^2}{dt^2}\left(\frac{g_i}{n}\right)\,=\,\frac{1}{2n}\partial_i\,g_k^2.
\label{3n}
\ee

Furthermore by inserting (\ref{3l}) into the Lagrangian (\ref{3c}) we obtain
\be 
L\,=\,\frac{n_0}{4\pi G}\,\int\,d^3\xi\,\frac{d}{dt}\left(\frac{g_i(\vec x(\vec \xi,t),t)}{n}\right)\,\dot x_i(\vec \xi,t)\,+\,\frac{1}{8\pi G}\int\,d^3x\,g_i^2(\vec x,t),
\label{3o}\ee
where we have used the identity
$$\int \,d^3\xi\,J\,g_i^2(\vec x(\vec \xi,t),t)\,=\,\int \,d^3x\,g_i^2(\vec x,t).$$

It can be easily checked that the Euler-Lagrange EOMs following from (\ref{3o}) are identical with (\ref{3m}) and (\ref{3n}).

The EOMs do not contain the gravitational constant $G$. The  Lagrangian contains it only as a common factor. Moreover, our modified Lagrangian (\ref{3o}) does not split into two parts involving a free and an interacting term. Clearly this strange property is a consequence of the phase space reduction due to  (\ref{3l}). We know only one other physical system possessing a similar property: the interacting Chaplygin gas if one eliminates the particle density from the Lagrangian (see \cite{d}, Sect. 2.1, item (i)).

\subsection{Relabeling symmetry and the transport equation for the gravitational field}

Any fluid dynamics in the Lagrangian formulation has to be invariant under  infinitesimal relabeling transformations
(volume preserving diffeomorphisms) (see \cite{c})
 \be
\vec \xi\quad \rightarrow\quad \vec \xi\,+\,\vec \alpha(\vec \xi),\quad \hbox{with}\quad \vec \nabla\cdot \vec\alpha\,=\,0.
\label{3p}\ee
                                                                                                                                                                     For the corresponding Noether charge  $Q$        we obtain  from the Lagrangian (\ref{3o}) 
\be
Q\,=\,\int\, d^3\xi \,\alpha_i(\vec \xi)\,\left(\frac{d}{dt}\left(\frac{g_k}{n}\right)\frac{\partial x_k}{\partial \xi_i}\,-\,\frac{g_k}{n}\frac{\partial \dot x_k}{\partial \xi_i}\right),
\label{3q}
\ee
where we have alerady performed one partial integration. But from the EOMs (\ref{3m}), (\ref{3n}) it follows that the integrand in (\ref{3q}) is already conserved {\it i.e.} that 
\be \theta_i\,\equiv\,\frac{\partial \dot x_k}{\partial \xi_i}\,\frac{g_k}{n}\,-\,\frac{\partial x_k}{\partial \xi_i}\,\frac{d}{dt}\left( \frac{g_k}{n}\right)
\label{3r}
\ee                                                                                                                                                                    
is a conserved quantity and so is a function of   $\vec\xi$         only. By solving (\ref{3r}) for   $\frac{d}{dt}\left(\frac{g_k}{n}\right)$             we obtain the once-integrated form of the EOM (\ref{3n})
 \be
\frac{d}{dt}\left(\frac{g_i}{n}\right)\,=\,\frac{g_k}{n}\,\frac{\partial \dot x_k}{\partial x_i}\,-\,\theta_k(\vec\xi)\frac{\partial \xi_k}{\partial x_i},
\label{3s}\ee                                                                                                                                                                      
which is a transport equation for the gravitational  field (cf. \cite{f} for the standard self-gravitating dust fluid).

\subsection{ Eulerian formulation}

In the Lagrangian formulation the fluid particles are labeled by comoving coordinates  $\vec \xi$. So a Lagrangian observer moves with the particle. On the other hand an Eulerian observer is located at a fixed position  $\vec x$   (see \cite{c}). Therefore we obtain the fluid fields in the Eulerian formulation from the ones given in the Lagrangian formulation by the following change of independent variables
 \be
(\vec \xi,t)\quad \rightarrow\quad (\vec x,t)\quad \hbox{with}\quad \vec x\,=\,\vec x(\vec \xi,t).\label{3t}
\ee                                                                                              
Correspondingly, we have to transform the time derivatives
\be
\frac{d}{dt}\quad \rightarrow\quad D_t\,=\,\frac{\partial}{\partial t}\,+\,u_k(\vec x,t)\frac{\partial}{\partial x_k}  \quad \hbox{ (convective}\quad \hbox{ derivative)}                     
\label{3u}\ee
with
\be
u_k(\vec x,t)\,\equiv\,\frac{d}{dt}\,x_k(\vec \xi,t)\vert_{\vec \xi=\vec\xi(\vec x,t)}.
\label{3v}
\ee

The EOMs then become:

{\bf Euler eq.}  
\be
D_t\,u_i\,=\,g_i.
\label{3w}
\ee
                                                                                                                                                   
{\bf Transport eq. for the gravitational field}:
\be 
D_t\left( \frac{g_i}{n}\right)\,=\,\frac{g_k}{n}\,\partial_iu_k\,-\,\theta_i,
\label{3y}
\ee
where the auxiliary variable  
$$\theta_i(\vec x,t)\,\equiv\,\frac{\partial \xi_k}{\partial x_i}\,\theta_k(\vec \xi)\vert_{\vec \xi=\vec \xi(\vec x,t)}
$$               obeys the EOM
\be 
D_t\,\theta_i\,+\,\theta_k\,\partial_i\,u_k\,=\,0
\label{3z}
\ee
and the particle density  $n$       obeys the 
{\bf continuity eq.}:                                          
\be 
\partial_t n\,+\,\partial_k(nu_k)\,=\,0.
\label{3za}
\ee

All these equations can be derived from the Lagrangian
\be
L\,=\,-\frac{1}{4\pi G}\int d^3x\left(g_i(D_t u_i-\frac{1}{2}g_i)\,+\,\theta(\partial_t n+\partial_k(nu_k))\,
-n\alpha D_t\beta\right).
\label{3zb}
\ee

Here   $\theta$, $\alpha$ and $\beta$     give the Clebsch-parameterization of the vector field  $\theta_i(\vec x,t)$ 
 \be
\theta_i\,=\,\partial_i\,\theta\,+\,\alpha\partial_i\beta.
\label{3zc}
\ee

They obey the Euler-Lagrange EOMs
\be
D_t\theta\,=\,D_t\alpha\,=\,D_t\beta\,=\,0.
\label{3zd}
\ee                                                                                                                                                                     
Operating with  $D_t$      on (\ref{3zc}) leads then to the EOM (\ref{3z}) for $\theta_i$. All the remaining EOMs are Euler-Lagrange EOMs derived from (\ref{3zb}).

\subsection{Coupling with baryonic matter}

We describe baryonic matter ($B$) by a standard self-gravitating fluid coupled to the darkon fluid by the jointly created gravitational field   $\phi$. Then, within the Lagrangian formulation, we have to add to (\ref{3c}) the following baryonic Lagrangian
  \be
L_B\,=\,m\,\int\,d^3\xi\,\left(\frac{(\dot x_i^B(\vec \xi,t))^2}{2}\,-\,\phi(\vec x^B(\vec \xi,t))\right),
\label{33a}
\ee  
                                                                                                                                                         where $m$ is the (constant) baryonic mass density in $\vec \xi$ -space.

Varying  $x_i^B$  leads to the Newton equation
\be
\ddot x_i^B(\vec \xi,t)\,=\,-\partial_i\,\phi(\vec x^B(\vec \xi,t),t),
\label{33b}
\ee                                                                                                                                                       whereas varying  $\phi$           in the total Lagrangian
\be
L\,=\,L_D\,+\,L_B\,+\,L_{field},
\label{33c}
\ee
                                                                                                                                                   where  $L_D$        is the darkon fluid Lagrangian (first term in (\ref{3e})), leads to the Poisson equation
  \be
\nabla \phi\,=\,4\pi G\,\left(\partial_k(nq_k)\,+\,\rho_B\right)
\label{33d}
\ee                                                                                                                                                                 
 whose r.h.s. consists of the sum of the darkon fluid source and the usual $B$-source. The $B$-mass density $\rho_B$       is defined by
  \be
\rho_B(\vec x,t)\,=\,m\,J^{-1}\vert_{\vec\xi=\vec \xi(\vec x,t)},
\label{33h}
\ee                           where $J$        is the Jacobian defined by (\ref{3h}).
We note that   $\rho_B$        satisfies the local conservation law (continuity equation)
 \be
\partial_t \rho_B\,+\,\partial_k(u_k^B\rho_B)\,=\,0
\label{33v}
\ee                                                                                                                                                                   
where   $\vec u^B$           is defined analogously to (\ref{3v}).

Adding baryonic matter leads to the following changes in the foregoing subsections:
The gravitational force appearing in $L_D$ is the total one    $g_i=-\partial_i\phi$                            and so is it in the case of the EOMs (\ref{3d}) and (\ref{3e}). The solution of the Poisson equation (\ref{33d}) is a superposition of the gravitational potentials created by the $B$ and the  $D$-sources
  \be
\phi\,=\,\phi_B\,+\,\phi_D
\label{new6}
\ee                                                                                                                                                               
with                                                                                                                                                   
\be
\triangle \phi_B\,=\,4\pi G \rho_B
\label{new7}
\ee                                                                                                                                                         
and
\be
\partial_k\,\phi_D\,=\,4\pi G n q_k.
\label{new8}
\ee
                                                                                                                                                                        
So (\ref{new8}) is now the substitute for (\ref{3l}) and, therefore, the conserved quantity            (\ref{3r}) becomes (note that $L_B$ is separately invariant w.r.t. relabeling transformations (\ref{3p}))
\be
\theta_i\,=\,-\frac{\partial{\dot x_k}}{\partial \xi_i}\frac{\partial_k\phi_D}{n}\,+\,\frac{\partial x_k}{\partial \xi_i}\,\frac{d}{dt}\left(\frac{\partial_k \phi_D}{n}\right).
\label{new9}
\ee
                                                                                                                                                                       
Then, in the Eulerian formulation, we obtain finally instead of (\ref{3y}) the transport equation   
\be
D_t\left( \frac{\partial_i\phi_D}{n}\right)\,=\,\frac{\partial_k \phi_D}{n}\,\partial_i u_k\,+\,\theta_i,
\label{new10}
\ee                                                                                                                                                                       
where $\theta_i$      obeys the EOM (\ref{3z}).

\section{   Hamiltonian  dynamics}

The derivation of the  Hamiltonian dynamics, {\it ie} the formulation of classical dynamics in terms of a Hamiltonian and a PB-algebra,  is an indispensable prerequisite for any quantization procedure.
We start with the Lagrangian (\ref{3zb}) from which we obtain immediately the Hamiltonian
\be
H\,=\,\frac{1}{4\pi G}\,\int\,d^3x\,\left(g_i(u_k\partial_k u_i-\frac{1}{2}g_i)\,-\,n\theta_iu_i\right),
\label{4a}
\ee                                  where we have used the identity \cite{d}
\be
\theta\,\partial_k(nu_k)\,-\,nu_k\alpha \partial_k\beta\,=\,-n\theta_ku_k\,+\,\partial_k(u_kn\theta)
\label{4b}\ee                                                                                                                                                                     
and then neglected the boundary contribution arising from the last term in (\ref{4b}).
For the Hamiltonian EOM
\be
\dot A(\vec x,t)\,=\,[A(\vec x,t),\,H]
\label{4c}
\ee                                                                                                                                                                     
we choose  as independent field variables $A\in(n,\,\theta,\,n\alpha,\,\beta,\,u_i,\,g_i)$.   
To read off from (\ref{4a}) and (\ref{4c}) the PBs for the $A$-fields we next rewrite their EOM as follows :                
 $$ \dot n\,=\,-\partial_k(u_kn),\quad \dot \theta\,=\,-u_k\partial_k\theta,$$                                                                                                                                                                   
$$ (n\alpha)^{\cdot}\,=\,-\partial_k(u_k\,n\alpha),\quad \dot \beta\,=\,-u_k\,\partial_k\beta,$$
\be
\dot u_i\,=\,-u_k\,\partial_k u_i\,+\,g_i\quad \hbox{and}
\label{4d}
\ee
 $$ \dot g_i\,=\,g_k\,\partial_iu_k\,-\,\partial_k(u_kg_i)\,-\,n\theta_i,$$                                                     
                                                                              where
$$\theta_i\,=\partial_i\theta\,+\,\alpha\partial_i\beta.$$

The EOM (\ref{4d}) have the Hamiltonian form (\ref{4c}) if the fluid fields $A$       obey the following nonvanishing PBs
\be
[\theta(\vec x,t),\,n(\vec x^{'},t)]\,=\,4\,\pi\,G\,\delta(\vec x-\vec x^{'})
\label{4e}
\ee
\be                                                                                                                                                                    
[\beta(\vec x,t),\,(n\alpha)(\vec x^{'},t)]\,=\,4\,\pi\,G\,\delta(\vec x-\vec x^{'})
\label{4f}
\ee                                                                                                                                                                        
and
\be 
[g_i(\vec x,t),\,u_j(\vec x^{'},t)]\,=\,4\pi \,G\,\delta_{ij}\,\delta(\vec x-\vec x^{'}).
\label{4g}
\ee                                                                                                                                                                                                                                                                                                                                                
For these PBs all Jacobi identities are trivially fulfilled. 
It is interesting to note that the PB-algebra (\ref{4e}-\ref{4g}) decomposes into three irreducible parts. Each component has a canonical structure which allows quantization by applying Dirac$'$s rule (details will be given in a separate paper)
\be
[A,\,B]_{PB}\,\rightarrow\,\frac{1}{i\hbar}\,[\hat A,\,\hat B],
\label{4h}
\ee                                                                                                                                                                         
where  $\hat A(\hat B)$                  denotes the quantum operator corresponding to the classical function  $A(B)$                    .
The PB-algebra (\ref{4e}-\ref{4g}) possesses a Casimir  (being a quantity whose PBs with all elements of the PB-algebra vanish) defined by
\be
C\,\equiv\,\int \,d^3x\,\epsilon_{ikl}\,\frac{\partial \theta_l(\vec x,t)}{\partial x_k}\,\theta_i(\vec x,t).
\label{4i}
\ee

To prove that $C$ is a Casimir we proceed in two steps:
\begin{itemize}
\item We choose a particular relabeling transformation (\ref{3p}) with $\alpha_i(\vec \xi)=-\epsilon_{ikl}\frac{\partial \theta_l(\vec \xi)}{\partial \xi_k}$.
The corresponding Noether charge           (\ref{3q}) is then given by
 \be
Q\,=\,\int\,d^3\xi\,\epsilon_{ikl}\frac{\partial \theta_l(\vec \xi)}{\partial \xi_k}\theta_i(\vec \xi).
\label{4j}\ee                                                                                                                                                          

But Eulerian variables do not respond to a relabeling of the Lagrange variable $\vec \xi$
\cite{d}. So $Q$       is a Casimir of the PB-algebra (\ref{4e}-\ref{4g}). 

\item It remains to express   $Q$           in terms of Eulerian variables.
Do do this we  get from the definition of $\theta_i(\vec x,t)$
\be
\theta_i(\vec \xi)\,=\,\theta_k(\vec x(\vec \xi,t),t)\,\frac{\partial x_k}{\partial \xi_i}.
\label{4k}
\ee
However, using (\ref{4k}) and the identity
\be
\epsilon_{ikl}\,\frac{\partial x_n}{\partial \xi_l}\,\frac{\partial x_m}{\partial \xi_i}\,=\,J\epsilon_{nmj}
\frac{\partial \xi_k}{\partial x_j}
\label{4l}
\ee
we obtain 
$$ \epsilon_{ikl}\frac{\partial \theta_l(\vec \xi)}{\partial \xi_k}\,\theta_i(\vec \xi)
\,=\,J\,\epsilon_{mjn}\frac{\partial \theta_m(\vec x,t)}{\partial x_j}\,\theta_m(\vec x,t)\vert_{\vec x=\vec x(\vec \xi,t)}$$
Hence we finally have
\be
Q\,=\,C.
\label{4m}
\ee
\end{itemize}

Of course the Casimir property of $C$ can be shown also directly by bracketing  (\ref{4i}) with all fluid fields and using the PB-algebra (\ref{4e}-{\ref{4g}).

We recall that $C$ vanishes if $\theta_i(\vec x,t)$ is given by the Clebsch-parametrization (\ref{3zc}) with non-singular functions $\theta,\alpha,\beta$ vanishing at infinity \cite{d}.

We note that the vortex helicity
\be
H_e\,\equiv\,\int\,d^3x\,\vec u(\vec x,t)\cdot \vec \omega(\vec x,t),
\label{4n}
\ee                                                                                                                                                          
where 
$$\vec \omega\equiv\vec \nabla\times \vec u$$
which is a Casimir in standard fluid dynamics, is not a Casimir in the case of our model. $H_e$ is a conserved quantity if  $\vec g$    is the gradient of a potential,  but it has a non- trivial PB with $\vec g$, namely:
\be
[g_i(\vec x,t),\,H_e]\,=\,8\pi \,G\,\omega_i(\vec x,t).
\label{4p}
\ee

Up to now we have used, within the Hamiltonian framework, those fluid fields which lead to the simplest possible form (\ref{4e}-\ref{4g}) for the PB-algebra. But in many applications it is more convenient to introduce instead of the auxiliary field  $\theta_i$        the momentum field $p_i$
\be
 4\pi G\,p_i(\vec x,t)\,=\,D_t\left(\frac{g_i}{n}\right)\,=\,\frac{g_k}{n}\,\partial_iu_k\,-\,\theta_i,
\label{4q}
\ee
where we have used the definition of $p_i$ in the Lagrangian formulation  
$$ p_i(\vec \xi,t)\,\equiv\,-\dot q_i(\vec \xi,t)\,=\,\frac{1}{4\pi G}\,\frac{d}{dt}\left( \frac{g_i}{n}\right)\vert_{\vec x=\vec x(\vec \xi,t)}
$$
and then changed to the Eulerian formulation.
In terms of the momentum density  $np_i$            we obtain for our PB-algebra, besides the PB-relation (\ref{4g}), 
\be
[(np_i)(\vec x,t),\,n(\vec x^{'},t)]\,=\,-n(\vec x,t)\,\partial_i\,\delta(\vec x-\vec x^{'}),
\label{4r}
\ee                                                                                           
    \be
[(np_i)(\vec x,t),\,u_k(\vec x^{'},t)]\,=\,\partial_i\,u_k\delta(\vec x-\vec x^{'}),
\label{4s}
\ee                                                                                                                                                                    

 \be
[(np_i)(\vec x,t),\,g_k(\vec x^{'},t)]\,=\,-g_k(\vec x,t)\,\partial_i\,\delta(\vec x-\vec x^{'}),
\label{4t}
\ee
         \be
[(np_i)(\vec x,t),\,(np_j)(\vec x^{'},t)]\,=\,-(np_j)(\vec x,t)\,\partial_i\,\delta(\vec x-\vec x^{'})+(np_i)(\vec x^{'},t)\,\partial'_j\,\delta(\vec x-\vec x^{'}),
\label{4u}
\ee                                                                                                                                                                                                                                                                                                                 which can be obtained from the transformation formulae 
 \be
n(\vec x,t)\,=\,\int\,d^3\xi\,\delta(\vec x-\vec x(\vec \xi,t)) 
\label{4v}
\ee
\be 
(nA)(\vec x,t)\,=\,\int\,d^3\xi\,A(\vec \xi,t)\,\delta(\vec x-\vec x(\vec \xi,t)),\quad A\in(p_i,u_i,g_i)
\label{4w}
\ee                                                                                                                                                                      
  and the canonical PBs in the Lagrange formulation (continuum generalization of (\ref{2q}))
 \be
[x_i(\vec \xi,t),\,p_j(\vec\xi^{\,\,'},t)]\,=\,\delta_{ij}\delta(\vec \xi-\vec \xi^{\,\,'})
\label{4y}
\ee                                                                                                                                                                        
and 
\be 
[\dot x_i(\vec \xi,t),\,q_j(\vec\xi^{\,\,'},t)]\,=\,\delta_{ij}\delta(\vec \xi-\vec \xi^{\,\,'})
\label{4z}
\ee                                                                                                                                                                        
                    if, in addition, we use the relation (\ref{3l}) for   $g_j$.

\section{ Space-time symmetries}

In this section we will show
\begin{itemize}

\item The darkon fluid possesses the unextended (zero mass) Galilei algebra $G_0$ as a symmetry algebra. The corresponding energy-momentum tensor is not symmetric  and contains a negative pressure term.
\item $G_0$ may be enlarged by dilations with dynamical exponent $z=\frac{5}{3}$. This particular value of $z$ is determined by the minimal coupling to gravity as given by eq. (\ref{2w}).
\item Expansions are not a symmetry of the darkon fluid.
\end{itemize}
 
\subsection{ Space-time translations and the energy-momentum tensor}

The linear momentum (generator of space translations) is given by
\be
P_i\,=\,\int\,d^3x \,(np_i)(\vec x,t),
\label{5a}
\ee                                                                                                                                                                           
where the momentum density  $np_i$             obeys the local conservation law
\be
\partial_t\,(np_i)\,=\,-\partial_k\,(u_k\,n\,p_i)\,+\,\frac{1}{8\pi G}\,\partial_i g_k^2\,=\,-\partial_k\,T_{ik}.
\label{5b}
\ee
Thus the  momentum  tensor
\be
T_{ik}\,=\,n\,p_i\,u_k\,+\,{\cal P}\delta_{ik}\quad          \hbox{with}\quad {\cal P}\,=\,-\frac{1}{8\pi G}\,g_k^2,
\label{5c}
\ee                                                   
is not symmetric and the pressure $\cal{P}$          is negative.

The Hamiltonian  (\ref{4a}) (generator of time translations), expressed in terms of the momentum density, takes the form
 \be
H\,=\,\int\,d^3x\,T_{tt}(\vec x,t)
\label{5d} 
\ee                                                                                                                                                                       
with
 \be
T_{tt}\,=\,n\,u_i\,p_i\,-\,\frac{1}{8\pi G}g_i^2.
\label{5e}
\ee                                                                                                                                                                       
The energy density   $T_{tt}$, of course,             obeys the local conservation law
 \be 
\partial_t T_{tt}\,=\,-\partial_k\,T_{tk}
\label{5f} \ee                                                                                                                                                                      
with   
$$T_{tk}\,=\,u_k\left( T_{tt}\,+\,{\cal P}\right).$$
                                                                                                                                                                          
If we identify, as usual, the  space-time component of the energy-momentum tensor with the momentum density                                                                                                                                                                    
 \be
T_{it}\,=\,np_i
\label{5h}\ee                                                                                                                                                                                                                                          
we may conclude, that the energy-momentum tensor $T_{\alpha,\beta}$ with $\alpha,\beta\in(t,k)$     obeys the standard conservation law
 \be
\partial_t\,T_{\beta,t}\,+\,\partial_k\,T_{\beta,k}\,=\,0.
\label{5i}
\ee                                                                                                                                                                      

But, because  $T_{\alpha \beta}$  is not symmetric, we note that         
\be
\partial_t\,T_{t\beta}\,+\,\partial_k\,T_{k\beta}\,\ne\,0.
\label{5j}
\ee
 
We remark that such a nonsymmetric energy-momentum tensor was introduced for the first time in 1947 by Weyssenhoff and Raabe \cite{fa}.                                                                                                                                                                      

\subsection{Zero mass Galilean symmetry}

Galilean boosts are generated by
\be 
K_i\,=\,tP_i\,-\,\frac{1}{4\pi G}\,\int \,d^3x\,g_i,
\label{5l}
\ee                                                                                                                                                                       
which is a conserved quantity because the EOM for the gravitational field  $g_i$       is given, due to (\ref{4d}) and (\ref{4q}), by
 \be
\dot g_i\,=\,np_i\,-\,\partial_k(u_kg_i).
\label{5la}
\ee 
                                                                                                                                                                     
The PB of the boost generator  $K_i$          and the momentum   $P_j$           vanishes
 \be [K_i,\,P_j]\,=\,0
\label{5m}
\ee                                                                                                                                                                      
as is easily see from the fact that $P_i$ is the generator of space translations    
 \be
[P_i,\,A(\vec x,t)]\,=\,\partial_i\,A(\vec x,t)\quad \hbox{for}\quad \hbox{any}\quad A.
\label{5n}
\ee                                                                                                                                                                      
The vanishing of the right hand side of (\ref{5m}) tells us that the total mass of the darkon fluid is zero.

\subsection{  Angular momentum}

The continuum generalization of expression (\ref{2na}) for the one-particle angular momentum is given by
\be
L_i\,=\,\epsilon_{ikl}\,n_0\,\int\,d^3\xi(x_k(\vec \xi,t)p_l(\vec \xi,t)\,+\,\dot x_k(\vec \xi,t)\,q_l(\vec \xi,t)).
\label{5o}
\ee                                                                                                                                                                        

By using the identity (\ref{3l}) and by passing to the Eulerian variables we obtain
\be
L_i\,=\,\epsilon_{ikl}\,\int\,d^3x\,\left(x_k\,n\,p_l\,-\,\frac{1}{4\pi G}\,u_k\,g_l\right).
\label{5p}
\ee
                                                                                                                                                                        
\subsection{Anisotropic scaling}

Anisotropic scaling, characterized by a dynamical exponent $z\ne1$, implies that the time and space variables transform differently  w.r.t. dilations
\be
t\,\rightarrow\,t^{*}\,=\,\lambda t,\quad \vec x\,\rightarrow \vec x^{*}\,=\,\lambda^{\frac{1}{z}}\vec x,\quad \hbox{with}\quad \lambda>0.
\label{5q}
\ee
                                                                                                                                                                     
In section 2.1 we have shown that the dynamics of free massless particles allows for an implementation of dilations with an arbitrary $z$.  To see what happens in the case of a darkon fluid we consider a continuum generalization of (\ref{2r}) of the generator of dilations $D$ 
 \be
D\,=\,tH\,-\,\frac{1}{z}\int\,d^3x\,np_ix_i\,-\,\frac{(1-\frac{1}{z})}{4\pi G}\,\int\,d^3x\,u_ig_i,
\label{5r}
\ee                                                                                                                                                                    
where in the last term we have used the relation (\ref{3l}) between the field   $\vec q$       and the gravitational field  $\vec g$. Taking now the time derivative of $D$ (\ref{5r}) and by using the EOMs (\ref{4d}) we obtain
\be
\frac{d}{dt}D\,=\,\frac{1}{8\pi G}\left(-3\,+\,\frac{5}{z}\right)\,\int\,d^3x\,g_i^2.
\label{5s}
\ee                                                                                                                                                                     

Thus $D$ is conserved if $z$ takes the value  $z=\frac{5}{3}$. It may easily be checked that with  $z=\frac{5}{3}$ all PBs  between $D$ and the elements of $G_0$ are in agreement with the PB-relations (\ref{2qa}).

The dilations are implemented by the following transformation of the fluid fields $A$
 \be
A(\vec x,t)\,\rightarrow\,A^{*}(\vec x,t)\,=\,\lambda^{\frac{z_A}{z}}\,A(\vec x^{*},t^{*}),
\label{5t}\ee                                                                                                                                                                                                                                                                                                                       where $z_A$        denotes the scale dimension of  $A$.           .
For an infinitesimal dilation we get
\be
[A(\vec x,t),\,D]\,=\,\left(t\partial_t\,+\,\frac{1}{z}(x_k\partial_k+z_A)\right)A(\vec x,t),
\label{5u}\ee                                                                                                                                                                      
which leads, due to the explicit form of $D$ (\ref{5r}) and the PBs (\ref{4g}), (\ref{4r}-\ref{4t}), to the following values for the $z_A$:
\be
z_n\,=\,3,\quad z_{\theta_i}\,=\,1,\quad z_{g_i}\,=\,\frac{7}{3},\quad z_{u_i}\,=\,\frac{2}{3}
\label{5v}
\ee
and therefore, due to (\ref{4q}), $z_{p_i}=1.$

Let us finally prove that our EOMs are not invariant with respect to expansions.
We prove this by demonstrating the contradiction.
So consider the expansions:
\be
t\,\rightarrow\,t^{*}\,=\,\Omega(t)t,\quad \vec x\,\rightarrow \,\vec x^{*}\,=\,\Omega(t)^{\frac{2}{z}}\vec x\quad \hbox{with}\quad \Omega(t)\,=\,\frac{1}{1-\kappa t}.
\label{5w}
\ee

Suppose that we would be able to implement such expansions. Then there would exist a generator $K$ of expansions which obeys, in combination with $H$ and $D$ an $O(2,1)$ algebra. On the other hand, as is well known \cite{a}, the PB-algebra between the elements of $G_0$ and the $O(2,1)$ algebra would not close for  $z\ne 2$. The PB between $K$ and the generator of boosts $K_i$ would then  imply the existence of a conserved generator of accelerations $F_i$
\be
[K,\,K_i]\,=\,\left(1-\frac{2}{z}\right)F_i.
\label{5y}
\ee 
                                                                                                                                                                       
But we know from the discussion given in section 3 that the relation (\ref{3l}) between the field  $\vec q$           and the gravitational field  $\vec g$     destroys the invariance of our EOMs w.r.t. arbitrary time dependent translations (gauge transformations), in particular w.r.t. to constant accelerations.  This contradicts the initial supposition of the existence of expansions as a symmetry of the EOMs. 

Note that the coupling of our darkon fluid to baryonic matter (see 3.4) will break dilation symmetry.

\section{ Darkon fluid cosmology}

We consider the darkon fluid as a model for the dark sector of the Universe. Hence we look for solutions of the fluid EOMs satisfying the cosmological principle (the Universe is supposed to be isotropic and homogeneous on large scales). The relevant cosmological EOMs are listed in subsection 6.1. In subsection 6.2 we derive and discuss the corresponding Hamiltonian dynamics. We start with the three cosmological EOMs in 1st-order form, derive the two constants of motion and construct out of them that Hamiltonian which has the correct scaling dimension.  Unique expressions for the corresponding PBs are obtained if we consider explicitly the implementation of dilation symmetry.  For that purpose it is convenient to introduce new independent variables with either scale dimension zero or one.  The PB-algebra possesses one Casimir which can be used to obtain a foliation of the three-dimensional phase space into an infinite number of two-dimensional phase spaces.  Within the two-dimensional phase space we get a completely integrable system. We are able to construct a corresponding Lagrangian.

Due to the appearance of a constraint it is not possible to quantize the arising Hamiltonian by applying the Dirac quantization rule.  An alternative way seems to be the affine quantization (see \cite{ha}).

In subsection 6.3 we solve the cosmological EOMs with the cosmological redshift $z$ as the independent variable.  By choosing some integration constants to be positive our model shows qualitatively the transition from a deceleration to an acceleration phase of the Universe. Fixing two integration constants by observational values we obtain a prediction for the cosmological history in general agreement with the existing data of the Hubble parameter.

\subsection{ Cosmological EOMs}

The realization of the cosmological principle within the Lagrangian formulation leads to the following structure for the field variables
\be 
x_i(\vec \xi,t)\,=\, a(t) \xi_i,
\label{9a}
\ee
\be 
n(\vec x,t)\,=\,n(t),
\label{9b}
\ee                                                                               
where  $ a(t)$   is the cosmological scale factor,
and       
\be   
g_i(\vec x,t)\,=\,\frac{n(t)}{n_0} g(a(t))x_i,\quad n_0=n(0).\label{9c}
\ee                                                                                                                                                  
The split of the r.h.s. of (\ref{9c}) into two factors and defining $ g$ as a function of $a(t)$ are a matter of convenience.
If we next put all this into the original EOMs  we obtain the following cosmological EOMs
\begin{itemize}
 \item     From the definition of  $ n$  (see (\ref{3g}))  we get
  \be               n(t)\,=\,\frac{n_0}{a^3(t)}.\label{9d}
\ee

\item    From the EOM   (\ref{3m}) and using   (\ref{9d})  we get
\be
\ddot a(t)\,=\,\frac{g(a)}{a^2(t)},\label{9e}
\ee                                                                                                                                                                

which is a Friedmann-like equation.
\item   From the EOM for     $g_i$  (\ref{3n})   and the definition of   $\theta_i$  (see (\ref{3r}))     we get
                                                                                                                                                               
\be
\dot g(a(t))\,=\,\frac{K_1}{a^2(t)},\quad K_1=\hbox{const.},\label{9f}
\ee

where  $K_1$      is related to    $\theta_i$       by
\be
\theta_i(\vec \xi)\,=\,-\frac{K_1}{n_0}\,\xi_i.\label{9g}
\ee
 \end{itemize}                                                                                                                                                        
We note that possible derivative terms in the gravitational coupling (\ref{2za}) give no contribution to the cosmological EOMs. Taking into account, 
additionally, baryonic matter (see 3.4) will not change the cosmological EOMs. In such a case $g(a)$ would be changed by an additive constant (see \cite{a15}).

\subsection{ Hamiltonian dynamics} 

The cosmological EOMs (\ref{9e}) and (\ref{9f}), when written in 1$^{st}$-order form, take the form
                                                                                                                                                              
     \be
\dot a\,=\,b,\quad \dot b \,=\, \frac{g}{a^2},\quad \dot g\,=\,\frac{K_1}{a^2}.\label{9h}
\ee

So we have to consider a dynamical system in a three-dimensional space. This finite dimensional space, called usually mini-superspace, will be the phase space for our Hamiltonian dynamics.
The dynamical system (\ref{9h}) possesses two constants of motion                                                                                                                                                  
    \be
 Q_2\,=\,bK_1\,-\,\frac{g^2}{2}\label{9i}
\ee
and                                                                                                                                           
  \be
Q_3\,=\,\frac{g^3}{6}\,+\,Q_2g\,+\,\frac{K_1^2}{a}\label{9j}
\ee

In order to construct Hamiltonian dynamics for the dynamical system (\ref{9h}) we have to find a Hamiltonian $H$ and the Poisson-brackets (PBs)     $[x_{\alpha},x_{\beta}]$                              for         $x_{\alpha}\in(a,b,g)$                                such, 
that the EOMs (\ref{9h}) become the Hamiltonian EOMs
                                                                                                                                                                
\be
\dot x_{\alpha}\,=\,[x_{\alpha},\,H]\,\equiv \,f_{\alpha},\label{9k}
\ee
where the PBs are required  to fulfill the Jacobi-identities and $H$ has to  be a function of the two constants of motion   $Q_2$       and     $Q_3$       only.
To reduce the arbitrariness in the construction of $H$ we note that the EOMs (\ref{9h}) are dilation invariant, {\it i.e.}  if    $x_{\alpha}$           are solutions of (\ref{9h}) then, by the transformation
$x_{\alpha}\rightarrow x_{\alpha}^{\star}$
\be 
x_{\alpha}^{\star}(t)\,=\,\lambda^{z_{\alpha}}\,x_{\alpha}(\lambda t),\quad \lambda> 0\label{9l}
\ee                                                                                                                                                       
  with $z_{a}=-\frac{3}{5}$, $z_b=\frac{2}{5}$,  $z_g=\frac{1}{5}$ we obtain another set of solutions.
For the scaling exponents of the constants of motion we obtain 
                                                                                                                                                                
\be
 z_{Q_2}\,=\,\frac{2}{5},\quad \hbox{and}\quad z_{Q_3}\,=\,\frac{3}{5}.\label{9m}
\ee
Let    $D$      be the conserved generator of dilations. Then we get from (\ref{9k}) and (\ref{9l}) that
                                                                                                                                                                 
\be
[x_{\alpha},\,D]\,=\,f_{\alpha}t\,+\,z_{\alpha}x_{\alpha}\,\equiv\,\eta_{\alpha}\label{9n}
\ee
(with no summation over   $\alpha$   in the middle term !)  leading to
                                 \be 
D\,=\,Ht-f,\quad \hbox{with}\quad \dot  f=H,\label{9o}
\ee
which is equivalent to                                                                                                                          
\be 
[H,\,D]\,=\,H.\label{9p}
\ee
However, this expression tells us that we have $z_{H}=1.$     Thus  neither    $Q_2$     nor   $Q_3$       but    $Q_2Q_3$     
    is a possible    candidate for the Hamiltonian $H$. So we choose
                                                                   \be
H\,=\,\frac{Q_2Q_3}{K_1^3}.\label{9q}
\ee
where the factor   $K_1^{-3}$         has been inserted in order to get a simple form for $H$ 
which will be obtained when we rewrite it in terms of some new variables which will be  introduced below.
With this $H$  the PBs for the    $x_{\alpha}$         have, according to (\ref{9k}), to be chosen such, that 
                             \be
[x_{\alpha},\,x_{\beta}]\,\,\frac{\partial H}{\partial x_{\beta}}\,=\,f_{\alpha}\label{9r}
\ee
holds. But (\ref{9r}) is equivalent to
                                                                                                                                                                 
\be
A_{\alpha\beta}\,y_{\beta}\,=\,f_{\alpha}\label{9s}
\ee
where
\be
y_{\beta}\,\equiv\,\frac{1}{2}\epsilon_{\beta \gamma\delta}\,[x_{\gamma},x_{\delta}]\quad \hbox{and}\quad A_{\beta\alpha}\,\equiv\,-\epsilon_{\beta\alpha\gamma}\,\frac{\partial H}{\partial x_{\gamma}}.\label{9t}
\ee

 The matrix $A_{\alpha\beta}$        is not invertible (it is anti-symmetric and of odd order). 
Therefore (\ref{9s}) has no unique solution for the   variables $y_{\beta}$.  Even if we require in addition, the validity of the Jacobi identities, the solutions for the  $y_{\beta}$     will not be unique.
 To render the PBs unique we must consider also the implementation of the dilation symmetry. 
First of all (\ref{9n}), (\ref{9o}) may be rewritten   as
\be
[f,\,x_{\alpha}]\,=\,z_{\alpha}x_{\alpha}\label{9u}
\ee                                                                                                                                                         and                                                                                                                                                           
\be
[f,\,D]\,=\,Ht.\label{9v}
\ee

From (\ref{9v}) we conclude that     $z_f=0$ . Thus $f$  is a function of the dilation invariant variables $x=\frac{ag^3}{K_1^2}$     and    $y=\frac{bK_1}{g^2}$  only. 
Then from     $z_{x}=z_y=0$                               and (\ref{9u}) we obtain
$$0\,=\,[f,\,x]\,=\,[y,\,x]\,\frac{\partial f}{\partial y}$$
and
$$0\,=\,[f,\,y]\,=\,[x,\,y]\,\frac{\partial f}{\partial x},$$
which shows  that the PB between $x$ and $y$ has to vanish                               \be 
[x,\,y]\,=\,0.\label{9w}
\ee
As the third variable (of our phase space)  besides $x$ and $y$ we choose  $w=\frac{g^2}{aK_1}$                   which has $z_w=1$.  Then, due to $z_H=1$ 
the Hamiltonian $H$ (\ref{9q}) has the structure            \be 
H\,=\,wh(x,y)\label{9y}
\ee
with                                                                                                                                                         
 \be
h(x,y)\,=\,\left(y-\frac{1}{2}\right)\left(1+x\left(y-\frac{1}{3}\right)\right).\label{9z}
\ee

In order to determine the remaining PBs  $[x,w]$                    and  $[y,w]$             we next rewrite the EOMs (\ref{9h})
in terms of the new variables $x$, $ y$  and $w$
\be
\dot x=w(xy+3),\quad \dot y=-\frac{2w}{x}\left(y-\frac{1}{2}\right)\quad \hbox{and}\quad \dot w=w^2\left(\frac{2}{x}-y\right).
\label{99a}\ee

But (\ref{99a}) are required to be EOMs derivable from the Hamiltonian $H$. So we get with (\ref{9y})
\be
[x,H]\,=\,[x,w]h\,=\,w(xy+3)\quad \hbox{and}\quad [y,H]\,=\,[y,w]h\,=\,-\frac{2w}{x}\left(y-\frac{1}{2}\right).
\label{99b}
\ee

Solving (\ref{99b}) for the two PBs gives us

\be
[x,\,w]\,=\,\frac{w(xy+3)}{h},\qquad [y,\,w]\,=\,-\frac{2w(y-\frac{1}{2})}{xh}.\label{99c}
\ee
It is easy to check that the PBs (\ref{9w}) and (\ref{99c}) satisfy all Jacobi-identities.

Expressing now the original variables $a$, $b$ and $g$ in terms of $ x$, $y$ and $w$
\be
a=K_1^{\frac{1}{5}}(xw)^{-\frac{3}{5}}x,\quad b= K_1^{\frac{1}{5}}(xw)^{\frac{2}{5}}y,\quad g= K_1^{\frac{3}{5}}(xw)^{\frac{1}{5}}\label{99d}
                          \ee
 we obtain their  PBs
\be
[a,b]\,=\,H^{-1}\left(\frac{2}{5}b^2+\frac{3}{5}\frac{g}{a}\right),\quad [a,g]\,=\,H^{-1}\left(\frac{bg}{5}+\frac{3}{5}\frac{K_1}{a}\right)\label{99e}
\ee and          $$[b,g]\,=\,-\frac{2}{5}\,\frac{K_1^3}{a^2Q_3}.$$

The equations (\ref{99e}) can be summarized by one equation
\be
[x_{\alpha},\,x_{\beta}]\,=\,H^{-1}(f_{\alpha}\,z_{\beta}\,x_{\beta}\,-\,f_{\beta}\,z_{\alpha}x_{\alpha}).\label{99f}
\ee                                                                                                                                                                      
(no summation over  $\alpha$     or $\beta$.)
The PB-algebra (\ref{99f})  is identical with a  construction given by Hojman \cite{g} who starts also with some EOMs, a Hamiltonian and a symmetry. But in contrast to \cite{g} our derivation  of (\ref{99f}) shows for the particular case of dilation symmetry the uniqueness of the PB-algebra.

\subsubsection{ One exact solution}
The EOMs (\ref{99a}) have a simple power law solution of the form                                                                                          
 \be
x(t)\,=\,-6,\quad y(t)\,=\,\frac{1}{2}, \quad\frac{1}{w(t)}\,=\,\frac{5}{6}(t-t_0)\label{99g}
\ee                                                                                                                                                                  
which for     $t_0=0$                          are dilation invariant. We will later show that (\ref{99g}) governs the behaviour  of the cosmological solutions at early times.
Furthermore the solution (\ref{99g}) determines a relative minimum of the Hamiltonian (\ref{9y}) at $H=0$. 

\subsubsection{ Reduced phase space}

In the following we will solve the EOMs (\ref{99a}) for $x$ as a function of $y$. So we reduce the dimension of phase space from three to two. To do that we can either use standard analytical methods or algebraic methods. The latter shows more clearly the physical nature of this reduction. For this reason we will present both methods.


{\bf Analytical method}

From the first two EOMs in (\ref{99a}) we obtain immediately an ODE of Bernoulli-type for $x(y)$

        \be 
x^{'}(y)\,=\,\frac{x(xy+3)}{1-2y},\label{99h}
\ee
where   $x^{'}(y)\,\equiv\,\frac{dx(y)}{dy}$ .  As usual we use the transformation
\be 
x(y)\quad \rightarrow\quad u(y)\,\equiv\,\frac{1}{x(y)}\label{99i}
\ee                     to obtain a linear ODE for $u(y)$
\be
u^{'}(y)\,+\,\frac{3u(y)}{1-2y}\,+\,\frac{y}{1-2y}\,=\,0.
\label{99j}\ee

The point    $y=\frac{1}{2}$                 is a singular point of the DE (\ref{99j}).Therefore we obtain one solution    $u_-(y)$       
for      $y<\frac{1}{2}$       and another one $u_+(y)$               for     $y>\frac{1}{2}$.                                     

\be
 u_-(y)\,=\,\frac{1}{3}\,-y\,+\,k_-\left(\frac{1}{2}-y\right)^{\frac{3}{2}},\qquad 
u_+(y)\,=\, \frac{1}{3}\,-y\,+\,k_+\left(y-\frac{1}{2}\right)^{\frac{3}{2}}\label{99k}\ee                    
      with real-valued constants   $k_+$ and $k_-$.          
Note that the first term in (\ref{99k}) represents a particular solution of the inhomogeneous  DE (\ref{99j}) whereas the second one represents the most general solution of the homogeneous part.
 The two solutions    $u_{\pm}$        are not connected to each other by an analytical continuation in $y$.  This implies that the reduced phase space consists of  two dynamically disconnected parts with either     $y<\frac{1}{2}$         or       with   $y>\frac{1}{2}$  .  We will show later that the validity of the condition               
 \be
y>\frac{1}{2}\label{99l}
\ee                                                                                                                                                                 
is  necessary in order to explain the accelerating expansion of the Universe. So we will use $y>\frac{1}{2}$  exclusively in the following.

 {\bf Algebraic method}

The PB-algebra (\ref{9w}), (\ref{99c}) possesses a Casimir
 \be
C\,=\,Q_3\,Q_2^{-\frac{3}{2}}\label{99m}
\ee       where we have used the equivalence between (\ref{99l}) and the condition 
                                                                                                                                                           
          \be Q_2\,>\,0.\label{99n}
\ee
To prove the Casimir-property of  $C$       we observe  first that    $z_C=0$. So  $C$        can depend only on $x$ and $ y$.  By using the definition of the  $Q_i$, $i=2,3$         and the transformation formula (\ref{99d}) we obtain
\be
C\,=\,(y-\frac{1}{2})^{-\frac{3}{2}}\left(\frac{1}{x}\,+\,y\,-\,\frac{1}{3}\right).\label{99o}
\ee
                                                                                                                                                                      
Therefore, due to (\ref{9w}), the PBs    $[x,C]$                  and    $[y,C]$      both vanish. To show   that $[w,C]=0$ too     we use (\ref{9w}) again by considering the chain of relations
\be
0\,=\,[H,\,C]\,=\,[w,\,C]\,\frac{\partial H}{\partial w}.\label{99p}
\ee
But as $\frac{\partial H}{\partial w}\ne0$             we see that $[w,C]$          must vanish.

This reasoning shows that any function of the independent variables which is a constant of motion and has a vanishing scaling exponent is a Casimir of our PB-algebra. But there exists no other function, functionally independent from (\ref{99m}), which possesses this property.
Now we use the fact that the equation
   \be C(x,y)\,=\,k,\qquad k\in \mathbb{R}^1,\label{99q}
\ee               defines, for different values of    $k$,       a foliation of the three-dimensional phase space (cp \cite{h}). So we may solve (\ref{99m}) for $x$ at fixed    $k$         and obtain
\be
x(k,y)\,=\,\frac{1}{\frac{1}{3}-y+k(y-\frac{1}{2})^{\frac{3}{2}}}, \qquad y>\frac{1}{2}
\label{99r}  \ee                                                                                                                                                                  
in accordance with (\ref{99k}) if we identify $k=k_+$.

This way we see that our system is described by a Hamiltonian dynamics on a two-dimensional phase space with variables $w$ and $y$ and  the Hamiltonian
\be
H\,=\,w\left(y-\frac{1}{2}\right)\left(1+x(k,y)\left(y-\frac{1}{3}\right)\right)\,=\,k\,w\,x(k,y)\,\left(y-\frac{1}{2}\right)^{\frac{5}{2}},\label{99s}
\ee                                                                                                                                                                     
where for the last equality  we have used (\ref{99r}), and
 the PB
 \be [w,\,y]\,=\,\frac{2w}{k\,x^2(k,y)\,({y-\frac{1}{2})^{\frac{3}{2}}}}\label{99t}
\ee                                                                                                                                                                       
leading together to the following Hamiltonian EOM
\be
\dot y\,=\,-\frac{2w}{x(k,y)}\left(y-\frac{1}{2}\right),\qquad \dot w\,=\,w^2\left(\frac{2}{x(k,y)}-y\right).\label{99u}
\ee

The dynamical system (\ref{99u}) is completely integrable. To show this we make use of the existence of a second constant of motion besides the Casimir  
 (see (\ref{9i}), (\ref{9j})). In this proof  we may use the Hamiltonian $H$ (\ref{99s}). Let us put $ H=E$, $E$=const. Then we can solve (\ref{99s}) for $ w$
\be
w(E,k,y)\,=\,\frac{E}{k\,x(k,y)\,(y-\frac{1}{2})^{\frac{5}{2}}}.\label{99v} \ee                                                                                                                                                                     
Inserting (\ref{99v}) into the first eq. in (\ref{99u}) we finally obtain $y(t)$ by a quadrature. 

{\bf Generator of dilations}

 We can also give an explicit expression for the generator of dilations in two-dimensional phase space.  From    $\dot f=H$              (\ref{9o}) we get $\dot y\frac{\partial f}{\partial y}=H$
and so
\be
\frac{\partial f}{\partial y}\,=\,-\frac{k}{2}\,x^2(k,y)\,\left(y-\frac{1}{2}\right)^{\frac{3}{2}}\equiv  F(y).
\label{99w}
\ee
                                                                                                                                                                       
{\bf Lagrangian}

The EOMs (\ref{99u}) are Euler-Lagrange EOM derived from the Lagrangian

                                                                                                                                                                        \be
L\,=\,\dot y\,F(y)\,\log w\,-\,H(w,y),\label{99y}
\ee
where   $H(w,y)$               is given by (\ref{99s}). To prove this it is sufficient to note that
                     \be \Pi(y,w)\,\equiv \,F(y)\,\log w\label{99z}
\ee
is the momentum canonically conjugate to $ y$. Indeed, by using (\ref{99t}) we obtain
\be
[y,\,\Pi(y,w)]\,=\,1.\label{999a}
\ee

 We can now use (\ref{99z}) to eliminate   $w$ from the Hamiltonian (\ref{99s}) to obtain 
\be
\label{9999b}                                                                                                                                                                          
H\,=\,k\,x(k,y)\,\left(y-\frac{1}{2}\right)^{\frac{5}{2}}\,\exp\left(\frac{\Pi}{F(y)}\right).
\ee
With (\ref{999a}) and (\ref{9999b}) we have finally Hamiltonian dynamics in a two-dimensional phase space expressed in terms of canonical variables    $y$      and  $\Pi$.

{\bf How to quantize this system?}

We would like to quantize the system (\ref{999a}) and (\ref{9999b}) with the constraint  (\ref{99l}). This cannot be done if we try to follow Dirac$'$s rule of going from the classical PB (\ref{999a}) to the quantum commutator
 \be
\label{9999c} 
[\hat y,\,\hat \Pi]\,=\,i\hbar.
\ee
                                                                                                                                                                         
The reason for this is that, for self-adjoint operators   $\hat y$             and    $\hat \Pi$          (\ref{9999c}) has, up to unitary equivalence, only one irreducible representation given by  \\
$\hat \Pi\,=\,\frac{\hbar}{i}\frac{\partial}{\partial y}$ ($-\infty<y<\infty)$ (cp \cite{ha}).
                  So the spectrum of  $\hat y$         covers the whole real axis.  To overcome this difficulty we follow J. Klauder  (see \cite{ha}) and consider instead of the canonical quantization (\ref{9999c}) the affine quantization given by $$[\hat q,\,\hat D]\,=\,i\hbar \hat q$$        with  $q\equiv y-\frac{1}{2}$                      and                                                         
\be
\label{9999d}
\hat D\,\equiv \frac{1}{2}(\hat \Pi\hat q\,+\,\hat q\hat \Pi)
\ee
and choose the irreducible representation for  $q>0$.
Details of such a quantization procedure still have to be worked out.

\subsection{Solution of the EOMs}


Let us first put the two constants of motion $Q_{2,3}$ equal to numbers $K_{2,3}$. Next we solve the defining equation of  $K_2$ (\ref{9i})    for   $\dot a=b$  and     
   obtain
\be 
\dot a \,=\,\frac{1}{K_1}\left( K_2\,+\,\frac{g^2}{2}\right).\label{999d}\ee

So, with    $K_2>0$,    equivalent to condition (\ref{99l}),             we must have
\be K_1>0\label{999e} \ee                     in order to describe an expanding Universe.

Using the definition of   $K_3$  (\ref{9j}) we obtain (we consider now $g$ as a function of the redshift $z$, \hfill \\ $(1+z)^{-1}\equiv a$),                                                                                                                                                                  
 a cubic equation  for $g(z)$:
\be
\frac{g^3}{6}\,+\,K_2 g\,=\,K_3\left(1-\frac{1+z}{1+z_t}\right)\label{999f}
\ee                                                                                                                                                                         
where the transition redshift  $z_t$             is defined by
\be
\frac{1}{1+z_t}\,=\,\frac{K_1^2}{K_3}\label{999g}.\ee

Let us look now at (\ref{999f}) for  $K_2>0$          and      $K_3>0$        together with the Friedmann-like equation (see (\ref{9e}))
 \be
\ddot a\,=\,g(z)\,(1+z)^2.\label{999h}
\ee                                                                                                                                                                        

For   $z>z_t$   we have  $g<0$          and so, due to (\ref{999h})   $\ddot a<0$. So, for  $z>z_t$                    we are in the deceleration phase of the early Universe.  On the other hand, for  $z<z_t$                   we have     $g>0$                 and then, due to (\ref{999h}),         $\ddot a>0$. So, for   $z<z_t$                      we are in the acceleration phase of the late Universe.
To get these results we have to require that  $K_{2,3}>0$.

To see this note that
\begin{itemize}
\item From (\ref{999g}) we conclude that $K_3>0$.
\item Let  $z_t$       be a zero of  $g$. Then differentiating  (\ref{999f})  at  $z=z_t$  shows that $\frac{K_2}{K_3}g^{'}(z_t)<0$.                  
But the transition from the decelerating to the accelerating phase requires  $g^{'}(z_t)<0$.                              .
Thus we must have   $\frac{K_2}{K_3}>0$. 
\end{itemize}
                       
Next, to simplify (\ref{999f}) we introduce
 \be
\tilde g(z)\,=\,K_2^{-\frac{1}{2}} g(z).\label{999i}\ee              Then (\ref{999f}) becomes (we recall that $k=K_3K_2^{-\frac{3}{2}}$)
\be
\frac{\tilde g^3}{6}\,+\,\tilde g\,=\,k\left( 1-\frac{1+z}{1+z_t}\right),\label{999j}\ee                                                                                                                                                                           
where  $\tilde g$             now  depends on the two parameters  $z_t$           and  $k$      only.

Using (\ref{999d}) the Hubble parameter $H(z)\equiv \frac{\dot a}{a}(z)$ may be expressed in terms of  $\tilde g$         
 \be
H(z)\,=\,\frac{K_2}{K_1}\left(1\,+\,\frac{\tilde g^2(z)}{2}\right)(1+z).\label{999k}
\ee                                                                                                                                                                          

However,  it is more convenient to consider instead of $H(z)$ the normalized Hubble parameter
\be
h(z)\,=\,\frac{H(z)}{H_0},\label{999l}
\ee                                                                                                                                                                           
for which we find
 \be
h(z)\,=\,\frac{(1+z)\left(1+\frac{\tilde g^2(z)}{2}\right)}{\left(1+\frac{\tilde g^2(0)}{2}\right)}\label{999m}
\ee

So, if we know  $H_0$, the cosmological history is determined in our model completely by the two parameters  $z_t$        and   $k$.                             

To obtain  $k$       in terms of observable quantities we consider the deceleration parameter
  \be
q(z)\,=\,-\frac{\ddot a}{aH^2}\,=\,\frac{h^{'}(z)}{h(z)} (1+z)\,-\,1,\label{999n}
\ee                                                                                                                                                                        
which, when expressed  in terms   of  $\tilde g$, $z_t$              and $k$ takes the form:
\be 
q(z)\,=\,-\frac{k(1+z)}{1+z_t}\,\tilde g(z)\left( 1+\frac{\tilde g^2(z)}{2}\right)^{-2}.\label{999o}
\ee
                                                                                                                                                                           
Thus the value of $k$          can be determined from the knowledge of the present deceleration parameter $q_0=q(0)$ and of $z_t$.

{\bf Remark}: The expressions for  $\tilde g(z)$, $h(z)$  and  $q(z)$   in our paper \cite{a15} can be obtained from the present ones by the substitutions $\tilde g(z)\rightarrow -\sqrt{6}\tilde g(z)$ and $k\rightarrow \sqrt{6} k$. 
Furthermore, the constants  $K_1$        and   $n_0$        are related to the constants   $D$         and $\beta$            from  \cite{a15}
by the relations  $K_1=-3D\beta$  and  $n_0\,=\,\frac{3D}{4\pi}.$

{\bf Asymptotic results} for   $z\gg z_t$               

From (\ref{999j}) we obtain 
\be
\tilde g(z)\,\sim\,-\left(6k\frac{z}{1+z_t}\right)^{\frac{1}{3}}. \label{999p}
\ee                                                                                                                           
Thus we obtain a power law behaviour for $h(z)$
\be
h(z)\,\propto \,z^{\frac{5}{3}},\label{999q}
\ee                                                                                                                                                                    
which shows that asymptotically our cosmological solutions are scale invariant.  On the other hand the deceleration parameter reaches asymptotically a constant value
\be
q(z)\sim \frac{2}{3}.\label{999r}
\ee

The asymptotic behaviour (\ref{999q}) differs from the predictions of the $\Lambda$CDM-model
\be
h_{\Lambda CDM}(z)\, \sim\,z^{\frac{3}{2}}
\label{extraa}
\ee
for the matter dominated epoch. But we expect a change of our prediction (\ref{999q}) if we generalise our model to the one that exhibits general covariance (cp. Section 9).

\subsection{Distances}

We must decide how to relate observational data of distances ({\it e.g.} of supernovae) 
 to the Hubble parameter $H(z)$. In standard FLRW cosmology we obtain, in terms of comoving coordinates for a flat Universe and for a radial light ray (see section 1.2 in \cite{R1}) the coordinate distance 
\be
\xi(z)\,=\,c \int_0^z\,\frac{dz'}{H(z')}.
\label{160}
\ee

In any relativistic theory of gravitation which possesses a metric the limit of an isotropic, homogeneous and flat Universe is described by the FLRW metric given in comoving coordinates by
 
\be
ds^2\,=\,a^2(t)\,(d\vec \xi)^2\,-\,c^2(dt)^2.
\label{161}
\ee                                                                                                                                                              
Then eq. (\ref{160}) arises from the requirement that light propagates along a null geodesic
\be
ds\,=\,0.
\label{162}
\ee                                                                                                                                                                 
Note that our darkon fluid model  is neither a relativistic nor a metric theory. However, it may be understood as a limiting case of such a theory as shown in section 9.
On the other hand, in a nonrelativistic theory we do not have an expanding space but only moving particles in a static space.  But both points of view lead to the same observable effects as has been already noted by Milne in 1934  \cite{b}. We need only to assume `{\it that, relative to a particular observer O the speed of light is a constant $c$ independent of the light source and that the classical formula for the Doppler effect is valid}' (McCrea, page 351 in \cite{R3}; see also section 5.2 in \cite{R4}). So both points of view allow us to use the expression (\ref{160}) for the coordinate distance in our model.

The dimensionless coordinate distance  $\xi_0$, defined by
\be
\xi_0(z)\,\equiv \,\frac{H_0}{c}\,\xi(z)\,=\,\int_0^z\,\frac{dz^{'}}{h(z^{'})}
\label{NN1}
\ee
may easily be computed in terms of the two parameters   $z_t$    and  $k$      by using (\ref{999m}) with (\ref{999j}) leading to (cp. Appendix A in \cite{a15})
\be
\xi_0(z)\,=\,6\left(1+\frac{\tilde g^2(0)}{2}\right) \, \sum_{i=1}^3\,a_i\,\log(x-x_i)\vert_{\tilde g(0)}^{\tilde g(z)},
\label{NN2}
\ee                                                                                                                                                                
where 
\be
a_i\,\equiv \, ((x_i-x_{i+1})(x_i-x_{i-1}))^{-1}
\label{NN3}
\ee                                                                                                                                                                
 $i  =  1,2,3$ and cyclic permutations and the  $x_i$  are the roots of the cubic equation:
\be
x^3\,+\,6(x-k)\,=\,0
\label{NN4}
\ee                                                                                                                                                                
given by
$$x_1\,=\,v_++v_-,\quad x_2\,=\,-\frac{v_++v_-}{2}\,+\,\frac{v_+-v_-}{2}i\sqrt{3},\quad x_3=x_2^{\star}
$$
with 
$$ v_{\pm}\,\equiv\,(3k\,\pm(8+(3k)^2)^{\frac{1}{2}})^{\frac{1}{3}}.$$

In terms of  $\xi_0$      we obtain, as usual, the dimensionless luminosity distance $D_L(z)$ 
\be 
D_L(z)\,\equiv\,(1+z)\xi_0(z),
\label{NN5}
\ee                 the related distance modulus $\mu(z)$
\be
\mu(z)\,\equiv\,5\log_{10}\,D_L(z)\,-\,5\log H_0\vert_{100kms^{-1}Mpc^{-1}}\,+\,42.38
\label{NN6}
\ee                       and the angular diameter distance  $D_A(z)$ 
 \be
D_A(z)\,=\,D_L(z)\,(1+z)^{-2}.
\label{NN7}
\ee                                                                                                                                                                 


\subsection{Predictions versus observations}

In subsection 6.3 we have shown already that our model explains qualitatively the transition from a deceleration phase of the early Universe to an acceleration phase of the late Universe if we choose the constants   $K_i (i = 1, 2, 3)$ to be positive. In this subsection we will show that
\begin{itemize}
\item  the quantitative predictions for the Hubble parameter $H(z)$ are in a general agreement with the observational data,
\item the distance moduli for GRBs seem to show a peak at large $z$ for the deceleration parameter q(z) in agreement with our predictions,
\item our unified model for the dark sector is unable to predict the present dark matter contribution $\Omega_{m0}$          and, therefore, it can neither explain the BAO data nor the data on the CMB-shift parameter by using the present models for them.
\end{itemize}

At present we do not know what the changes for the prediction of the Hubble parameter (\ref{999m}) would  look like had we generalized our model to a general covariant theory presented in section 9. Therefore we have not performed, so far, any `least squares fits’ to the data.

\subsubsection{ Hubble parameter}
In fig 1 we compare the best out of three predictions of our model for $h(z)$ given in \cite{a15} with observed Hubble parameters which we adopt from table 1 of [30]. In addition we show the $\Lambda$CDM-fit to these data taken from \cite{i}


\begin{figure}
    \begin{center}
\includegraphics[angle=0,width=0.8\textwidth]{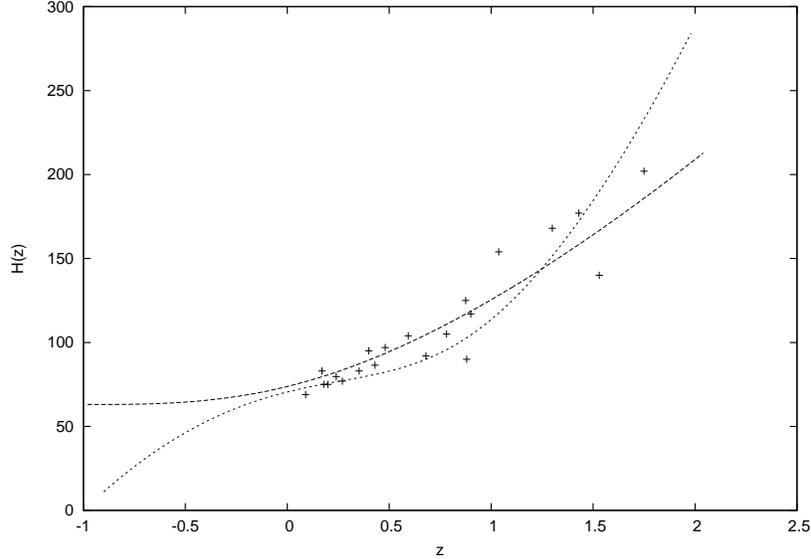}
			\end{center}
\caption{Comparison of experimental data from [32], \cite{l}, \cite{j} for  $H(z)$ measured in units of km s$^{-1}$ Mpc$^{-1}$ with our predictions for $z_t=0.71$, $q_0=-0.57$ and $H_0=70.5$ (dotted line) and a $\Lambda$CDM fit for $H_0=73.8$ and $\Omega_{m,0}=0.27$ taken from \cite{i} (dashed line).} 
\end{figure}
In \cite{a15}    we used
   $z_t=0.71$                  and    $q_0=-0.57$              taken from a   $\Lambda$CDM-     fit to some data given in \cite{m}. 
Let us point out that these values for   $z_t$              resp.  $q_0$     are in agreement with very recent model independent constraints on cosmological parameters  \cite{n}, \cite{o}, \cite{p}.  For our prediction we have fixed the normalization in fig. 1 by choosing  $H_0=70.5$km\,$s^{-1}$\,Mpc$^{-1}$   in accordance with the value used very recently in \cite{o}.
The corresponding values for the three constants $K_i$ turn out to be 
$$K_1\,=\,2.361\times 10^5 (km\, s^{-1}Mpc^{-1})^3\,=\,8.033\times 10^{-54}s^{-3}$$
$$K_2\,=\,1.263\times 10^7 (km\, s^{-1}Mpc^{-1})^4\,=\,1.392\times 10^{-71}s^{-4}$$
$$K_3\,=\,9.530\times 10^{10} (km\, s^{-1}Mpc^{-1})^6\,=\,1.103\times 10^{-106}s^{-6}$$
leading to $k=2.123$.

\subsubsection{Remarks on distance moduli data}

The most recent supernovae (SN Ia) distance moduli data are those of the Union 2.1 compilation \cite{RF4}. 
But we do not see any necessity to give a detailed comparison of these data with the predictions of our model (see eq. (195))
 because it has been claimed very recently by O. Farooq et al. \cite{RF2} that the recent 21 Hubble parameter versus redshift data points
 (see fig. 1) give constraints on DE models which are almost as restrictive as those from the Union 2.1 compilation.
In addition to the SN Ia data we have for larger $z$ values up to $z = 8.1$ distance moduli data from Gamma-Ray Bursts (GRBs)
 at hand (see table II in \cite{RF5}]).  Using such data one obtains the tentative result that the deceleration parameter $q(z)$ shows 
a maximum roughly at $z \sim 1.6$ \cite{RF6} in qualitative agreement with the predictions from our model (see fig. 2 in \cite{a15}).

\subsubsection{Remarks on BAOs and the CMB-shift parameter}

Baryon acoustic oscillations in the early Universe are imprinted as peaks in the late time matter power spectrum (cp. \cite{RF7}).  
But the physical interpretation, mainly of the first peak position, depends heavily on the present baryonic 
and dark matter contents   $\Omega_{b0}$       resp. $\Omega_{m0}$          of the Universe (see fig. 4 in \cite{RF7}. {\it This issue is usually ignored 
in the data analyses performed in the literature} \cite{RF8}. But our model, as a unified model for the dark sector of the Universe,
 is unable to predict  $\Omega_{m0}$         (for more details cp. Section 6.6) and therefore we cannot at present analyse these peaks.
The same argument holds for the CMB-shift parameter (cp. \cite{RF9}). This issue has been discussed already in our recent paper \cite{a15}.

\subsection{Comparison with alternative models for dark energy}

In the following we will sketch how to perform a comparison of our model with alternative models of the  dark energy. Any details will be left for the future.

The Friedmann equations, derived from GR including a cosmological `constant' $\Lambda$, are for a flat Universe (see {\it e.g.} \cite{R1})
\be
H^2\,=\,\frac{8\pi G \rho}{3}\,+\,\frac{\Lambda}{3},
\label{88a}
\ee

 \be
\frac{\ddot a}{a}\,=\,-\frac{4\pi G}{3}\,(\rho+3p)\,+\,\frac{\Lambda}{3},\label{88b}
\ee

where $\rho$     is the total energy density of the Universe (sum of the matter ($M$) and dynamical dark energy ($DE$) if we neglect radiation) and,  if we assume that matter is described by pressure less dust, $p$ is the (supposed negative) pressure of $DE$
\be
\rho\,=\,\rho_M\,+\,\rho_{DE},\quad p\,=\,p_{DE}\,<\,0.
\label{88c}
\ee                                                                                                                                                               

Let us now assume that the first Friedmann equation (\ref{88a}) holds. Then the second equation (\ref{88b}) can be replaced, as is well known, by the continuity equation
\be
\frac{d}{dt}\left(\frac{\Lambda}{8\pi G}\,+\,\rho\right)\,+\,3H\,(\rho+p)\,=\,0.
\label{88d}
\ee

Note that if we allow a weak time-dependence of Newton's gravitational constant $G$ the last equation (\ref{88d}) would have to be generalised (see \cite{R9}, \cite{R10}). However, this case will not be discussed here, instead we refer the reader to \cite{R10} and the literature cited therein.

At this stage it is important to contrast here two possibilities  for the description of the `dark energy' \cite{R9}:
\begin{itemize}

 \item  Some kind of fluid described by an equation of state (EoS) $W(z)$
                                                                                                                                                                 
\be W(z)\,=\,\frac{p}{\rho_{DE}}
\label{88e}
\ee

\item A (possibly) time-dependent $\Lambda$ with a fixed EoS
                                                                                                                                                                 \be
\rho_{\Lambda}\,\equiv\,\frac{\Lambda}{8\pi G}\,=\,-p_{\Lambda}.
\label{88f}
\ee
\end{itemize}
                                                                                        
Now our Friedmann-like equations are (see (\ref{999h}) and (\ref{999k}))

\be
\frac{\ddot a}{a}\,=\,K_2^{\frac{1}{2}}\,\tilde g(z)\,(1+z)^3,
\label{88g}
\ee
and
\be
H^2\,=\,\left(\frac{K_2}{K_1}\right)^2\left(1\,+\,\frac{\tilde g^2(z)}{2}
\right)^2(1+z)^2.
\label{88h}
\ee

Comparing the corresponding right hand sides with (\ref{88a},\ref{88b}) we obtain two equations for the effective functions  $\rho$, $\rho_{\Lambda}$      and p
                                                                                                                                                               \be
\rho+\rho_{\Lambda}\,=\,\frac{3H^2}{8\pi G}\,=\,\frac{3}{8\pi G}\,\left(\frac{K_2}{K_1}\right)^2\left( 1+\frac{\tilde g^2(z)}{2}\right)^2\,(1+z)^2,
\label{88i}
\ee
\be
p-\rho_{\Lambda}\,=\,-\frac{1}{8\pi G}\,\left(\frac{K_2}{K_1}\right)^2\,\left(1+\frac{\tilde g^2(z)}{2}\right)^2(1+z)^2\,-\,\frac{1}{4\pi G}K_2^{\frac{1}{2}}\,\tilde g(z)(1+z)^3.
\label{88j}
\ee

In the first scenario we have to put  $\rho_{\Lambda}=0$                  and arrive at two equations for the three functions  $\rho_M$, $\rho_{DE}$          and $p$. If we suppose that we have no interaction between $CDM$ and $DE$ we have, in addition, local conservation of $\rho_M$                               
     \be
\dot \rho_M\,+\,3H\rho_M\,=\,0
\label{88k}
\ee
with the solution
                                                                                                                                                                 \be
\rho_M\,=\,\rho_{M,0}(1+z)^3.
\label{88l}
\ee

From (\ref{88i}), (\ref{88j}) and (\ref{88l}) we easily get an equation for the effective EoS which, however, contains as a free parameter the present value of the energy density of matter ($CDM$ + baryonic matter),  $\rho_{M,0}$,  which is not determined in our unified model for the dark sector of the Universe, {\it i.e.} $\rho_{M,0}$         is a free parameter in our model (for a very detailed discussion of this issue see [47]). Therefore our model, as any other unified dark fluid model, is not able to predict an EoS for $DE$. However, on the other hand, we can define an EoS  $W_D$      for the darkon fluid as a whole
                                                                                                                                                                   \be
W_D\,=\,\frac{p}{\rho}\,=\,-\frac{1}{3}\,-\,\frac{2 K_2^{\frac{1}{2}}\tilde g(z)(1+z)}{3\left(\frac{K_2}{K_1}\right)^2\left(
1+\frac{\tilde g^2(z)}{2}\right)^2}.
\label{88m}
\ee

Whether   $W_D$      for some $z$-region shows a phantom-like behaviour ($ w_D  < - 1 $) will be left for future considerations.

In the second scenario we have to put  $\rho_{DE}=p=0$                          which corresponds to model II in \cite{R10}. We decompose  $\rho_M$         into its $DM$ and $B$ parts
\be
\rho_M\,=\,\rho_{DM}\,+\,\rho_B
\label{88n}
\ee
                                                                                                           and take into account the local conservation of  $\rho_B$  which leads, in the cosmological limit, to (cp. (\ref{88l}))
                                                                                                                                                                    \be
\rho_B\,=\,\rho_{B,0}(1+z)^3.
\label{88o}
\ee

Note that    $\rho_{B,0} $, in contradistinction to  $\rho_{M,0}$, is directly observable. Then (\ref{88i}), (\ref{88j}) lead to predictions for  $\rho_{DM}$        and   $\rho_{\Lambda}$     for a given set of constants $K_i$ ($i=1, 2, 3$).  More details and their comparison with the results communicated in \cite{R10} will be left for the future.

\section{Modeling dark matter halos by a steady state darkon fluid}

We start with the darkon fluid EOMs in the Eulerian formulation as given in section 3. By taking their time-independent form (steady state equations) we will derive a nonlinear ordinary differential equation (ODE) for the gravitational potential in the spherically symmetric case. The solutions of this ODE may serve as a model for the dark matter (DM) halos of galaxies. In particular we consider DM-dominated galaxies in which case the halo dominates the galactic gravitational potential and therefore determines the rotational velocity of stars (rotation curves (RCs)).

In subsection 7.1 we  derive the nonlinear ODE for the gravitational potential in the spherically symmetric case and the behaviour of their solutions for large distances from the centre of the halo. Then, in subsection 7.2, we present the results of the numerical solutions of this ODE and compare them with the main characteristics of the observational data for the RCs of DM-dominated galaxies.

\subsection{ Equation for the gravitational potential}

From the darkon fluid EOMs (\ref{4d}) we obtain immediately the following steady state equations:

Steady state equation for the gravitational field
\begin{equation}
u_k\partial_k\left(\frac{g_i}{n}\right)\,=\,\frac{g_k}{n}\partial_iu_k\,-\,\theta_i,
\label{7a}
\end{equation}
                            where the auxiliary field  $\theta_i$            is given by the solution of
\be
u_k\partial_k \theta_i\,+\,\theta_k\partial_i u_k\,=\,0.
\label{7b}\ee
                                                                                                                                                                    
Euler equation:
\be
u_k\partial_k u_i\,=\,g_i
\label{7c}
\ee
                                                                                                                                                                    
Continuity equation:
\be
\partial_k(u_k n)\,=\,0.
\label{7d}
\ee
                                                                                                                                                                    
Next we consider the halos as spherically symmetric objects. Then the vector fields appearing in (\ref{7a}-\ref{7d}) take the following form:
\be
u_k(\vec x)\,=\,\frac{x_k}{r}u(r),\quad g_k(\vec x)\,=\,-\frac{x_k}{r}\phi^{'}(r),
\label{7e}
\ee
$$\theta_k(\vec x)\,=\,\frac{x_k}{r}\,\theta(r),$$
                                      where $\phi$         is the gravitational potential.  By using (\ref{7e}) in the steady state equations we obtain from (\ref{7a})
      \be
u^2\left(\frac{\phi^{'}}{un}\right)^{'}\,=\,\theta,
\label{7f}
\ee                                                                                                                                                             
where   $\theta$, obtained by the integration of (\ref{7b}), is given by
      \be
\theta(r)\,=\,-\frac{1}{2}\frac{\gamma}{\alpha u(r)}.
\label{7g}
\ee

For convenience we have chosen to write the integration constant as  $-\frac{\gamma}{2\alpha}$. Euler's eq. (\ref{7c}) leads to the Bernoulli eq.
      \be
\frac{1}{2}u^2\,=\,-\phi
\label{7h}
\ee                                                                                                                                                            
and the solution of the continuity eq. gives
\be
un\,=\,\frac{\alpha}{r^2},
\label{7i}
\ee
      where $\alpha$       is an integration constant. 

As in the Lagrangian formulation the vector field  $\vec \theta$             is a conserved quantity we may wonder whether it is possible to connect the integration constants   $\alpha$      and  $\gamma$       appearing here and  $K_1$         appearing in the cosmological context (see (\ref{9g})).

 
We find, in our case,                                                                                                                                                                   
\be
\theta_k(\vec \xi)\,=\,-\frac{1}{2}\xi_k\vert \vec \xi\vert \,\frac{n_0\gamma}{\alpha^2}.
\label{new.1}
\ee


   

This expression has a different $\vec \xi$ dependence from that of the  r.h.s. of (\ref{9g}).  
So, unfortunately, there is no relation between $\alpha$ and $\gamma$ on one hand and $K_1$ on the other.  
                                                                                                                                                                 

 We note that the solution (\ref{7i}) produces a  singularity at the origin on the r.h.s. of the continuity eq. (\ref{7d}). This singularity might be interpreted as the central black hole of the galaxy (see \cite{s} for the existence of a central black hole in LSB-galaxies). But a black hole describes a sink for any kind of matter.  So we get    $\alpha<0$              and therefore 
\be
u\,<\,0.
\label{7j}\ee

Inserting all these expressions into (\ref{7f}) we obtain a ODE 
   \be
\frac{1}{r^2}(r^2\phi^{'})^{'}\,=\,\frac{\gamma}{2r^2}(-2\phi)^{-\frac{3}{2}}
\label{7k}
\ee                                                                                                                                                                 
which may be understood as the Poisson eq. for a self consistently determined effective DM mass density $\rho$
 \be
  4\pi G\, \rho(r)  \,=\,\frac{\gamma}{2r^2} (-2\phi)^{-\frac{3}{2}},
\label{7l}
\ee   
                                                                                                                                                           Next we switch over to a dimensionless potential  $\phi$     by the substitution      
\be
\phi\,\rightarrow \,\gamma^{\frac{2}{5}}\phi.
\label{new220}
\ee
  From the foregoing equations we easily conclude that   $\gamma$          has the dimension   $m^5s^{-5}$                                 as it should be.

 So, altogether, we end up with a nonlinear, non-autonomous ODE for $\phi$
 \be
(r^2\phi^{'})^{'}\,=\,\frac{1}{2} (-2\phi)^{-\frac{3}{2}}, 
\label{7m}
\ee                                                                                                                                                                 
in which $\phi$       is nonpositive and satisfies the natural boundary conditions
                                                        \be 
\phi(\infty)\,=\,0\quad  \hbox{and}\quad \phi(1)\,=\,-\frac{\eta}{2},\quad \eta>0.
\label{7n}
\ee

 We note that the ODE (\ref{7m}) and the first boundary condition    ($\phi(\infty)=0$)             are invariant w.r.t. to the scale transformation    
\be
\phi(r)\,\rightarrow \tilde \phi(r)\,\equiv \,\phi(\lambda r),\,\lambda>0.
\label{7o}
\ee                                                                                                                           
Thus the second boundary condition is necessary to make the solution unique.

To simplify (\ref{7m}) we introduce, instead of  $\phi$        the function  $\psi=-2r\phi$ which then has to  satisfy the ODE  
   \be
\psi^{''}\,=\,-r^{\frac{1}{2}}\,\psi^{-\frac{3}{2}}(r),
\label{7p}
\ee                                                                                                                                                                  
whose positive solutions have to be determined.  
                                                                                                                                
{\bf Remark}: Eq. (\ref{7p}) resembles the Thomas-Fermi equation which arises in the self-consistent determination of the potential within a neutral atom.
 But there are two important differences between these two problems. Besides the negative sign in (\ref{7p}), the function on its r.h.s. is reciprocal to the one appearing in the Thomas-Fermi equation. 

To determine the required solutions of (\ref{7p}) let us consider first (\ref{7p}) with the boundary conditions
\be
\psi(1)\,=\,\eta,\,\eta>0\quad \hbox{and}\quad \psi^{'}\,=\,\beta.
\label{7q}\ee
                                                                                                                                                                   
In this case we can use the following lemma, which is a particular case of a more general theorem derived by S. Taliaferro (see theorem 2.1 in \cite{t})

{\bf Lemma  \cite{t}}. Let  $\psi(r)$         be a positive solution of (\ref{7p}) satisfying (\ref{7q}) for fixed  $\eta>0$        and an arbitrary real number  $\beta$.
Then there exists  $r_2$, satisfying $1<r_2<\infty$                                 such that $\lim_{r\rightarrow r_2}\psi(r)=0$                           and   $\psi(r)$            cannot be continued past $r_2$           as a twice continuously differentiable solution of (\ref{7p}).

To use this lemma to the study of properties of required solutions of (\ref{7m}) it is advantageous to perform first the transformation
  \be
r\,\rightarrow\, x\,\equiv\,  \log r,\quad \psi(r)\,\rightarrow \,\varphi(x)\,\equiv\,e^{-x}\psi(e^x).
\label{7r}
\ee

Then (\ref{7p}) with the boundary conditions (\ref{7q}) gets transformed into the following autonomous ODE
 \be
\varphi^{''}(x)\,+\,\varphi^{'}(x)\,=\,-\varphi^{-\frac{3}{2}}(x)  
\label{7s}
\ee                                                                                                                                                               
with the boundary conditions  
\be
\varphi(0)\,=\,a, \,a>0\quad \hbox{and}\quad \varphi^{'}(0)\,=\,b,\,b\in R^1
\label{7t}\ee  
where $a$ and $b$ are dimensionless numbers.                                                                           
                                                                                                                                                        
By (\ref{7r}) the lemma may be translated to the following statement about $\varphi$: For fixed $a$     and arbitrary $b$ we have $\lim _{x\rightarrow x_2}\varphi(x)=0$, where   $x_2\equiv \log r_2$,        and    $\varphi$       cannot be continued past  $x_2$.
Eqs. (\ref{7s}) and (\ref{7t}) are together equivalent to the integral equation
\be
\varphi(x)\,=\,a\,+\,b\left(1\,-\,e^{-x}\right)\,-\,\int_0^{x}\,ds\,\left(1\,-\,e^{s-x}\right)\,\varphi^{-\frac{3}{2}}(s).\label{7u}\ee
                                                                                                                                                                  
Using the relation  
\be 
\phi(r)\,=\,-\frac{1}{2}\varphi(\log r)
\label{7v}
\ee                                       
       (\ref{7u}) may be rewritten as an integral equation for $\phi(r)$:

 $$ \phi(r)\,=\,-\frac{1}{2r}\,\int_0^r\,ds(-2\phi(s))^{-\frac{3}{2}}\,-\,\frac{1}{2}\int_r^{\infty}\frac{ds}{s} (-2\phi(s))^{-\frac{3}{2}}$$
\be
+\,\frac{1}{2}\left( -a -b \,+ \,\int_1^{\infty}\frac{ds}{s} (-2\phi(s))^{-\frac{3}{2}}\right)  
\label{7w}
\ee
$$
+\,  \frac{1}{2r}\left( b \,+ \,\int_0^{1}\,ds (-2\phi(s))^{-\frac{3}{2}}  \right),$$                                                                                                                                                         
which is cast in such a form that the first line contains the usual integral of the Poisson equation, the second line a constant and the third one a term being proportional to $\frac{1}{r}$.

 But the latter is in conflict with the original ODE (\ref{7m}). The reason is that the transformation (\ref{7v})  becomes singular at $r = 0$. 
Therefore the coefficient of $\frac{1}{r}$ has to vanish which fixes the value of $b$ at
\be
b\,=\,b_c\,\equiv\,-\int_0^1ds (-2\phi(s))^{-\frac{3}{2}}\,=\,-\int_{-\infty}^0ds \,e^s\,\varphi^{-\frac{3}{2}}(s)\label{7y}
\ee
                                                                     
 But the lemma given above holds for any   $b$, and so in particular also for    $b=b_c$. Thus, due to the relation      (\ref{7v}),    it holds also for   $\phi(r)$
             {\it  ie.} $ \lim _{r\rightarrow r_2}\phi(r)=0$                      and so $\phi(r)$         cannot be continued past   $r_2$.

Next we return to looking at solutions   $\varphi$         of (\ref{7s}).  We know from the results given above, that
\begin{itemize}
\item we have to find a positive solution  $ \varphi$       at least for   $0<x<x_2$    with   $\varphi(x_2)=0$.                 
\item we have  to vary $b$ until it reaches a value  $b_c$   which corresponds to
\be
b_c\,=\,-\int_{-\infty}^0\,ds\,e^s\,\varphi^{-\frac{3}{2}}(s)\label{7z}
\ee                                                                                                                                                                
   leading to
\be
\varphi^{'}\le\,-e^{-x}\int_{-\infty}^x ds\,e^{s-x}\,\varphi^{-\frac{3}{2}}(s)\,<\,0\quad \hbox{for}\quad \hbox{all} \quad b\le b_c.\label{zaa}\ee
\end{itemize}

To relate the solutions          of (\ref{7s}) to the RCs  $v(r)$           of a DM-dominated galaxy we recall that  $v(r)$  is determined from the 
equality of the centripetal acceleration and the gravitational force acting on a star in circular motion \cite{u}     
                                                                                                                                                                          
\be 
\frac{v^2(r)}{r}\,=\,\gamma^{\frac{2}{5}}\phi^{'}(r),
\label{7ab}
\ee
where the factor $\gamma^{\frac{2}{5}}$ arises due to the dimensionlessness of $\phi$ after the substitution (\ref{new220}).
Then, using (\ref{7v}) we can determine $v$           in terms of the derivative of   $\varphi$
\be
v^2(r)\,=\,-\frac{1}{2}\gamma^{\frac{2}{5}}\varphi^{'}(x)\vert_{x=\log r}.
\label{7ac}
\ee

\subsection{ Predictions versus observations}

To begin with  we would like to discuss how to fix, at least in principle, scales for distances and velocities.  Due to the dilation symmetry of our model  (see (\ref{7o})) we have no intrinsic scale for distances.  A scale for velocities may be chosen by fixing the value of the dimensionfull  constant  $\gamma$.  However, $\gamma$,  being an integration constant,  is an extrinsic parameter. 
So our model  does not possess any  intrinsic scale for distances  or for velocities.  To get  intrinsic scales we must introduce dilation symmetry breaking terms into the Lagrangian. We can do this either by taking into account the coupling to baryonic matter and/or by adding within the gravitational coupling (\ref{2w}) higher order derivative terms as indicated in (\ref{2za}).

So, for the present form of our model, we get the required scales by fixing two extrinsic parameters.        For velocities we have to fix  $\gamma$. For distances we have to fix the scale parameter  $\lambda$       in (\ref{7o}) or, equivalently, put $ x = 0$  in (\ref{7t}) which fixes $ r = R$ where $R$  is the chosen unit of length. 
Thus, to fit observational data for the RCs of any given galaxy, we would  have to perform a best fit for          $\gamma$, $R$ and the dimensionless `initial' value  $a$    in (\ref{7t}).  However,  as we view  the present model  only as  a first building block for a new theory (cp. section 9) we think it is premature to  perform  any `least square fits'  to the data.  Thus we have  only  checked whether there is  a general qualitative agreement of the predictions of our model with the data.  We have done this  by fitting the extrinsic parameters $\gamma$ and $R$ to one particular galaxy at a fixed value of $a$.
 Note that the choice of   $a$       fixes $ b_c$. 

 More detailed discussion has been postponed  until we have generalised our model to a general covariant one as suggested in section 9.


\begin{figure}
   \begin{center}
\includegraphics[angle=0,width=5cm]{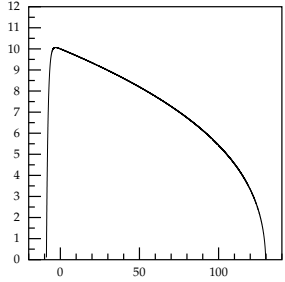}
			\end{center}
\caption{Numerical solution           of (\ref{7s}) for $a = 10$ and $b = -0.03$  } 
\end{figure}
We have solved numerically the nonlinear differential equation (\ref{7s}) for  $\varphi(x)$           by choosing the initial conditions (\ref{7t}) with $a = 10$ and $b \le -0.03$. The main result is the abrupt change of the character of the graphs for  $\varphi(x)$          at negative $x$-values at the critical $b$-value $b_c$ given approximately by 
                        \be
b_c = -0.031475314466,   
\label{7ad}
\ee
                                                                                                                                                                  
For $b > b_c$ the graph of $\varphi(x)$ crosses the $x$-axis at some $x_0 < 0$ where $x_0$ is an increasing function of $b$. We believe, although this is numerically hard to verify, that         $x_0\rightarrow-\infty$                   for  $b$ decreasing to $b_c$. For $x < x_0$ the solution becomes complex valued. This behaviour is illustrated in fig.2 ($b = -0.03$) and fig.3 ($b = b_c$).

\begin{figure}
  \begin{center}
\includegraphics[angle=270,width=5cm]{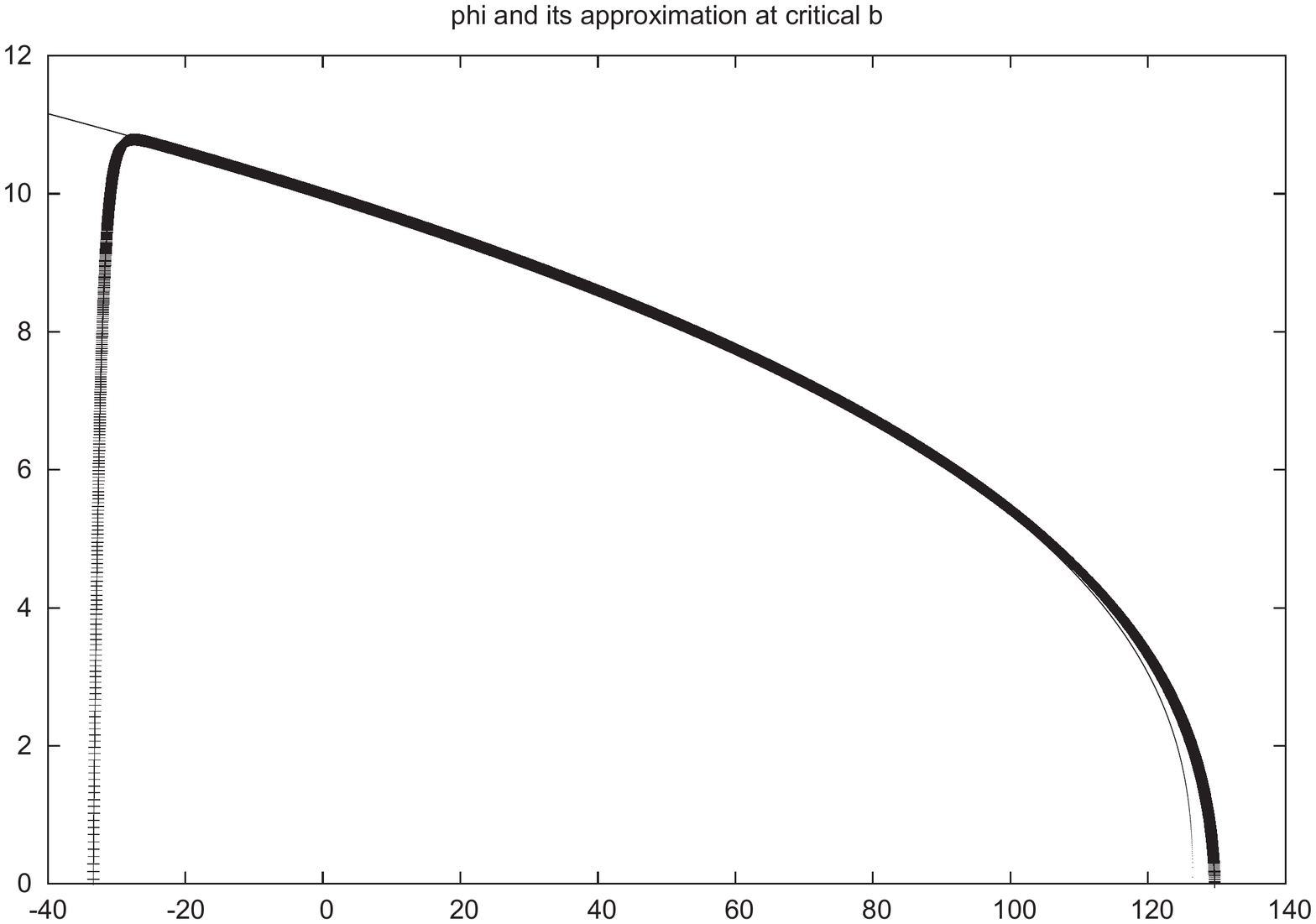}
			\end{center}
\caption{Numerical solution       of (\ref{7s}) for $a = 10$ and $b = b_c$ (thick line) 
compared with our approximation (\ref{extra}) (thin line)} 
\end{figure}

On the other hand for $b < b_c$ the graph of  $\varphi(x)$        starts to rise rapidly at some $x < 0$ and goes to infinity. This behaviour is clear from fig.4 ($b = -0.032$).

\begin{figure}
    \begin{center}
\includegraphics[angle=0,width=5cm]{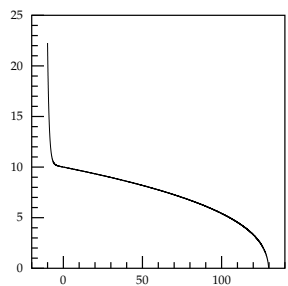}
			\end{center}
\caption{Numerical solution              of (\ref{7s}) for $a = 10$ and $b = -0.032$ } 
\end{figure}

Furthermore, from the plot of  $\varphi^{'}(x)$              for $b = -0.032$ (fig. 5) we see that  $\varphi^{'}(x)<0$ for all $x\in R^1$.

\begin{figure}
    \begin{center}
\includegraphics[angle=0,width=5cm]{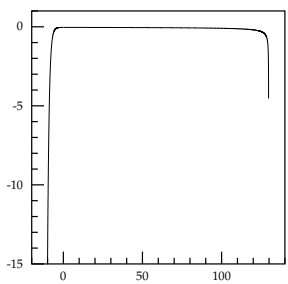}
			\end{center}
\caption{Numerically obtained  $\varphi^{'}$                  for $a = 10$ and $b = -0.032$} 
\end{figure}

The behaviour of the graphs of  $\varphi$          and  $\varphi^{'}$               for  $b\ge b_c$       resp.  $b<b_c$                     as shown in figs. $2-5$ strongly suggests that the $b_c$ obtained numerically in (\ref{7ad}) is, modulo numerical fine tuning,  indeed identical to the  expression for $b_c$ derived in section 7.1 as the correct $b$-value for modeling halos by means of the solution  of the boundary value problem (\ref{7s},\ref{7t}) for $a = 10$.

Hence we have to compare qualitatively the RCs calculated for $b = b_c$ with observational data. For that we plot in figs. 6-7 the rotational velocities $v(r)$ determined from eq. (241). In fig 6. 
we show the behaviour of $v(r)$ at small radii with the extrinsic parameters $\gamma^{\frac{1}{5}}$, resp. $R$ fitted, for fixed $a=10$ 
to the RC of the dwarf galaxy UGC 8490 (data taken from [56]). Fig 7. shows the global behaviour of $v(r)$ in dimensionless units.



\begin{figure}
    \begin{center}
\includegraphics[angle=0,width=8cm]{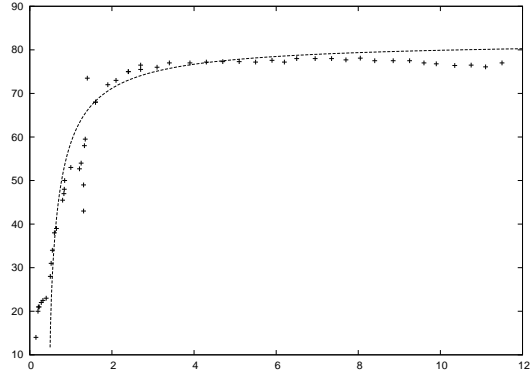}
			\end{center}
\caption{Numerically determined plot of the darkon fluid contribution to the rotational velocity $v(r)$ (in $km s^{-1}$) at small radii (in $kpc$) for $b = b_c$, $\gamma^{\frac{1}{5}}=615 kms^{-1}$ and $R=4\times10^{11} kpc$. Compared to the RC of UGC 8490 (data from [56])} 
\end{figure}

  \begin{figure}
  \begin{center}
\includegraphics[angle=0,width=5cm]{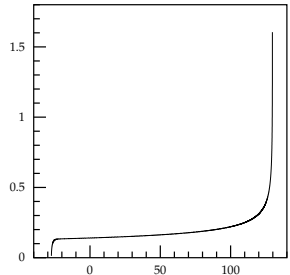}
			\end{center}
\caption{Global behaviour of the rotational velocity $v(r)$ for $b = b_c$ plotted as a function of $x=\log r$ } 
\end{figure}

The graphs for the darkon fluid contribution to the RCs given in figs. 6-7 show the following qualitative characteristics when compared to observational data on DM haloes (for a very recent review see \cite{r})
\begin{itemize}
\item They start at a radius $r_1 \sim 0$  and  rapidly increase reaching $ v(r) \sim  76 kms^{-1}$ at $r \sim 4kpc.$ (fig. 6).


\item Beyond this increase our model predicts a rather flat, but still very slowly increasing behaviour for the RCs until extremely large radii. This is basically in agreement with all observations according to which the RCs of a galaxy stay approximately constant {\it after attaining a maximum at about 5 kpc}  \cite{w}. This statement is true, in particular, for the DM-dominated galaxies: the LSB-galaxies (cp. the RCs reproduced in \cite{y} and in figs. 3,4 of \cite{z}) and dwarf galaxies (cp. \cite{ra} and the RCs reproduced in [56]).

The behaviour of $\varphi(x)$ within this flat-like region can be well approximated by (see fig.3)
\be
\varphi(x) \,\sim\,  f(x) = (2.5)^{0.4} (c-x)^{0.4},
\label{extra}
\ee                                                                                            
where $c  \sim    126.5$ has been fixed by $f(0) = 10$. Then we obtain $f'(0) \sim      -0.03162$, which is a good approximation for $b_c$ (\ref{7ad}).  $f(x)$ is the solution of the ODE
\be
f'(x)\,=\,-f^{-\frac{3}{2}}(x)
\label{extra1}
\ee                                                                                                                                    and so is an approximate solution of  the ODE (\ref{7s}) if $\varphi'\sim$      const.

\item They terminate at an extremely large radius roughly given by $r_2\sim e^{130}$ with a diverging RC as $r \rightarrow r_2 $.

 Of course, this contradicts all observations. But the size of $r_2$ is clearly orders of magnitude larger than the size of the visible part of the Universe:  In the FRW model the size $R_0$ of the Universe is given by $R_0 = 0.5 ct_0$ \cite{rb}, where $t_0 \sim 13.7$ Gyr is the age of the Universe \cite{u}. Thus we obtain $R_0\sim 2.1$ Gpc. 
The choice $R=4.10^{14}pc$ in fig. 6 gives  $r_2 \sim 10^{71}$ pc, which is  about 62 orders of magnitude larger than $R_0$. 
\end{itemize}

We conclude that this highly unphysical behaviour of $v(r)$ as $r\rightarrow r_2$ is possibly a consequence of considering only an isolated halo. 
 We expect an improvement of this situation if we embed the halo in the time-dependent darkon fluid (to be done by an appropriate generalization of the considerations in the next section 8).

\section{Influence of the cosmic expansion on binary systems}

In the following we consider two astrophysical objects, one of large mass $M$ ({\it e.g.} the sun) and another one of small mass $m \ll M$ ({\it e.g.} a planet) embedded in the time dependent darkon fluid
which, asymptotically behaves like the cosmological solution (see section 6.2). We would like to assess the change of Newton's force of attraction between the two bodies due to the presence of the darkon fluid. To be specific we approximate this binary system by a point-like test particle of mass $m$ which moves in the time-dependent gravitational potential created by a point-like mass $M$ and the overall darkon fluid considered in the rest system of the heavy mass $M$.

 Within GR the corresponding problem is to find, as a solution of Einstein's equation, an interpolation between the Schwarzschild and the FLRW metric. Then the motion of the test particle is the geodesic motion in this metric. A solution to this problem has been given  in a very recent paper \cite{R11} (see also the recent review \cite{R12} and the literature quoted therein, which covers also the history of this problem starting with the paper by Mc Vittie in 1933 \cite{R13}). In the Newtonian limit one obtains for the EOM of the test particle (see \cite{R11})
  \be
\ddot x_i\,=\,\frac{\ddot a}{a}x_i\,-\,\frac{GM}{r^3}x_i.
\label{eight.one}
\ee 
                                                                                                                                                                  
It turns out that the correction to Newton's law by the present accelerated expansion is completely negligible within the solar system (see section III A in \cite{R12}).

In the following we will sketch the ideas how to obtain the correction to Newton's law for our model.

We start with the EOMs given in 3.4 with a gravitational potential 
\be
\phi(r,t)\,=\,\phi_D(r,t)\,-\,\frac{GM}{r},
\label{eight.two}
\ee      where  $\phi_D$       represents the darkon fluid contribution. Then we obtain, in the spherically symmetric case, the EOMs
\be
\dot w\,+\,(uw)^{'}\,=\,0,
\label{eight.three}
\ee 

 \be
\dot u\,+\,\frac{1}{2}(u^2)^{'}\,=\,-\phi^{'},
\label{eight.four}
\ee 
                                                                                                                                                                    
 \be
(r^2\phi_D^{'})^{\cdot}\,+\,u(r^2\phi_D^{'})^{'} \,=\,w\theta,
\label{eight.five}
\ee 
 
\be
\dot \theta\,+\,(u\theta)^{'}\,=\,0,
\label{eight.six}
\ee                                                                                                                                                                                                                                                                  
where the functions  $u$          and    $\theta$    are given by (\ref{7e})
 generalized to the time-dependent case and we have defined  $w=r^2 n$.   A dot represents $\partial_t$           and a slash  $\partial_r$.
           .
Next we have to solve (\ref{eight.three}-\ref{eight.six}) with the following boundary conditions:
\begin{itemize}
\item for  $r\rightarrow \infty$          the functions  $w,u,\phi_D$                          and   $\theta$             attain their cosmological limits defined in 6.1
  \be
w(r,t)\,=\,\frac{r^2 n_0}{a^3(t)},\quad u(r,t)\,=\,r\frac{\dot a(t)}{a(t)},
\label{eight.seven}
\ee                           
$$\phi_D(r,t)\,=\,-\frac{r^2}{2}\,\frac{g(a(t))}{a^3(t)},\quad \theta(r,t)\,=\,-\frac{K_1}{n_0a^2(t)}r.$$

They are solutions of (\ref{eight.three}-\ref{eight.six}) if  $a$       and   $g$      satisfy the cosmological equations (\ref{9e}), (\ref{9f}).
\item    for  $r\rightarrow  0$ we require that
\be
r\phi_D\,\rightarrow\,0
\label{eight.eight}
\ee                                                                                                                                                             leading, if inserted into the EOMs (\ref{eight.three}-\ref{eight.six}), to the following behavior  of $u$, $w$, $\theta$       and      $\phi_D$       for
 $r\rightarrow  0$
\be
u(r,t)\,\sim\,\beta r^{-\frac{1}{2}}                                                                                                    \qquad \hbox{with}\quad                 \beta\,=\,\pm(2GM)^{\frac{1}{2}},
\label{eight.nine}
\ee
\be
w(r,t)\,\sim\,r^{\frac{1}{2}}\,f_w\left(t-\frac{2}{3\beta}r^{\frac{3}{2}}\right),
\label{eight.ten}
\ee
\be
\theta(r,t)\,\sim\,r^{\frac{1}{2}}\,f_{\theta}\left(t-\frac{2}{3\beta}r^{\frac{3}{2}} \right),\label{eight.eleven}
\ee
 \be
\phi{'}_D\,\sim\,\frac{2}{5\beta}\,r^{\frac{1}{2}}\,(f_wf_{\theta})\left(t-\frac{2}{3\beta}r^{\frac{3}{2}}\right),\label{eight.twelve} 
\ee                                                                                                                                                                                                                                                                                                                                               
where the functions   $f_w$, $f_{\theta}$             and the sign of $\beta$  still have to be determined.
\end{itemize}
If it turns out that  it is not be possible to obtain analytic expressions for these functions one should try to find an approximation involving  a two-scale perturbation expansion. The small spatial scale will be given by the size of the binary system and the large one by the cosmic scale. This problem is currently under consideration.                                                                                                                                                 


\section{Outlook for a general covariant theory}

The Poincare algebra treats spatial and time variables on the same footing. Then its enlargement by dilations has necessarily a dynamical exponent $z=1$ (cp. \cite{R5}). But in our model we have anisotropic scaling with $z=\frac{5}{3}$.  Therefore, an enlargement of the Milne gauge group (\ref{2t}) to arbitrary space-time diffeomorphisms ({\bf Diff}) as the symmetry group for a generalized darkon fluid model seems to be forbidden. But the dilation symmetry with $z=\frac{5}{3}$ was an emergent symmetry and  not an input of our model. So we should look for a general covariant generalization of our model without imposing any form of scale symmetry.

To do this  we have first to restore the Milne gauge invariance for the Eulerian action (\ref{3zb}) by undoing the gauge fixing (\ref{3l}) leading to
\be
S\,=\,\frac{1}{4\pi G}\int dt\,d^3x\,\left(4\pi Gnq_i(D_tu^i-g^i)\,-\,\theta(\partial_tn+\partial_k(nu^k))\,+\,n\alpha D_t\beta\right)\,+\,S_{field}
\label{nine1}
\ee                                                                                                                                                           with the spatial metric given by  $\eta_{ij}=\delta_{ij}$ and $S_{field}$ is determined by the expression (31).

Now we propose the following action as the {\bf Diff}-generalization of (\ref{nine1}) 
\be
D_{diff}\,=\,\frac{1}{4\pi G}\int d^4x\,\sqrt{-g}\left(4\pi Gn\tilde q_{\nu}u^{\lambda}\triangle_{\lambda}u^{\nu}\,-\,\theta\triangle _{\lambda}(nu^{\lambda})\,+\,n\alpha u^{\lambda}\partial_{\lambda}\beta\right)\,+\,S_{EH},
\label{nine2}
\ee
where we have used the following notation (in units with $c=1$):
$g$  is the determinant of the covariant space-time metric $g_{\mu\nu}$       with the invariant line element
 \be
ds^2\,=\,g_{\mu\nu}\,dx^{\mu}\,dx^{\nu} 
\label{nine3}
\ee
and     $\triangle_{\lambda}$  denotes the covariant derivative
\be
\triangle_{\lambda}u^{\nu}\equiv \partial_{\lambda}u^{\nu}\,+\,\Gamma_{\lambda\sigma}^{\nu}u^{\sigma}.
\label{nine4}\ee
Here $\Gamma_{\lambda \sigma}^{\nu}$    is the corresponding Levi-Civita connection, 
$n$                                                                                                                                                                                        is the scalar particle density and   $u^{\lambda}$      the four-velocity.   $\theta$, $ \alpha$       and   $\beta$      are scalar fields.  The vector field  $\tilde q_{\mu}$       will be deduced, as given below, from the Milne-scalar  $q_i$. Finally the last term in (\ref{nine2}), 
which is the {\bf Diff}-generalisation of the last term in (\ref{nine1}), is the Einstein-Hilbert action given by
\be
S_{EH}\,=\,-\frac{1}{16\pi G}\int d^4x\,\sqrt{-g}R,
\label{nine5}
\ee                                                                                                                                                                         
where $R$ is the Ricci-scalar.

To understand (\ref{nine2}) as a {\bf Diff}-generalization of (\ref{nine1}) we consider the limiting case
\be
u^0\rightarrow 1, \quad g_{\mu\nu}\rightarrow \eta_{\mu \nu}\qquad (\eta_{\mu\nu}\equiv \hbox{diag}(-1,1,1,1,))
\label{nine6}
\ee
$$ \Gamma_{\lambda\sigma}^{\nu}\rightarrow 0\quad \hbox{except}\quad \Gamma_{00}^i\rightarrow -g^i,$$
which are the leading terms known from the Newtonian weak field limit of General Relativity (GR)
$g_{\mu\nu}=\eta_{\mu\nu}+h_{\mu\nu},\,\vert h_{\mu\nu}\vert \ll1$   and an expansion of $h_{\mu\nu}$        in inverse powers of $c$. Here $g^i$ is the nonrelativistic gravitational field as used in (\ref{nine1}).

To construct the vector field  $\tilde q_{\mu}$         we proceed in two steps: 
\begin{itemize}
\item
By means of a Lorentz transformation we promote the Milne-scalar  $q_i$         ($i=1,2,3)$ to     $q_{\underline \alpha}$           ($\underline \alpha$         = 0,1,2,3), which is a scalar field w.r.t. {\bf Diff} but a contravariant space-like vector field, labeled by  ${\underline \alpha}$, in tangent space.
\item By means of tetrads  $E_{\mu}^{\underline \alpha}$         (known from teleparallel gravity, cp. \cite{RR8}) we transform  $q_{\underline \alpha}$         to the four-vector field $\tilde q_{\mu}$     
 \be
q_{\underline \alpha}\,\rightarrow \tilde q_{\mu}\equiv q_{\underline \alpha}E_{\mu}^{\underline \alpha}.
\label{nine7}\ee                                                                                                                                                                          
\end{itemize} 

                                                                                                                                                                 The tetrads   $E_{\mu}^{\underline \alpha}$    transform like a contravariant vector w.r.t. {\bf Diff} by changing the Greek index and like a four- vector w.r.t. Lorentz transformations by changing the underlined Greek index. The tetrads         are related to the metric tensor by
\be
g_{\mu\nu}\,=\,\eta_{\underline{\alpha}\underline{\beta}} \,E_{\mu}^{\underline{\alpha}}\,E_{\nu}^{\underline {\beta}}.
\label{nine8}
\ee                                                                                                                                                                           
So the limit  $g_{\mu\nu}\rightarrow\eta_{\mu\nu}$ corresponds to  $E_{\mu}^{\underline \alpha}\rightarrow \delta _{\mu}^{\underline \alpha}$   and so to  $\tilde q_{\mu}\rightarrow q_{\underline \mu}$                    .
Therefore we obtain,   in the limiting case, for the individual terms in (\ref{nine2})
$$
\tilde q_{\nu}u^{\lambda}\triangle_{\lambda}u^{\nu}\rightarrow q_i(D_tu^i-g^i),$$
\be
\triangle_{\lambda}(nu^{\lambda})\,\rightarrow \,\partial_tn+\partial_k(nu^k),\quad u^{\lambda} \partial_{\lambda}\beta\rightarrow D_t\beta
\label{nine9}
\ee                                                                                                                                           in agreement with (\ref{nine1}).

What about possible scaling properties of the action (\ref{nine2})? The generators of {\bf Diff} do not form a closed algebra together with the generators of anisotropic scaling transformations
\be
t\,\rightarrow\,t^{*}=\lambda^zt,\quad x_i\,\rightarrow\,x^{*i}=\lambda x^i.
\label{nine10}
\ee
                                                                                                                                                                           
A closed algebra can be obtained only if the rigid transformations (\ref{nine10}) are replaced by local ones (Weyl transformations). But the Einstein-Hilbert action (\ref{nine5}) does not respect the Weyl symmetry (for that we would have to replace (\ref{nine5}) by the Weyl action (cp. \cite{RR9} and the literature cited therein)). So we can only expect, like in Horava gravity \cite{R6}, to find scale invariant solutions in the ultraviolet or in the infrared regime.
Any further consequences of this new ansatz (\ref{nine2}) for a general covariant darkon fluid have still to be worked out. In particular we are interested in the cosmological implications. But due to the larger number of gauge fields in the action (instead of the three gravitational fields  $g^i$      we have now ten fields of the metric tensor  $g_{\mu\nu}$) we will obtain  cosmological EOMs other than the ones studied so far. In particular we will obtain a prediction for the Hubble parameter $h(z)$ which would probably be different from (\ref{999m}).  Therefore nothing can be said at the moment about the asymptotic behaviour of $h(z)$.

\section{Conclusions and outlook}

In the present paper we have reviewed our recent results on the darkon-fluid model which is, as outlined in section 9, a first building block for a new general covariant theory describing the dark sector of the Universe. The model involves new physics by using nonrelativistic massless particles with a nonstandard coupling to the gravitational field. The model, which contains no free parameters in its Lagrangian, predicts qualitatively  correct values of the
 late time cosmic acceleration as well as the flat behaviour of galactic rotation curves. These successes encourage us to continue our work. We have to work out the astrophysical implications of the covariant theory and have to compare them with the predictions of the present model and with observational data. 

So the main open topics of research are:

\begin{itemize}
\item Derive the EOMs of the covariant theory.
\item Restrict the EOMs to the cosmological regime and compare their predictions with the present model. Perform a best fit to the Hubble parameter resp. distance moduli data.
\item Enlarge the covariant model by adding baryonic matter.
\item Look for the solutions of the spherically symmetric but time dependent EOMs which attain for large distances their cosmological limits.
\item  Apply the results to
\begin{itemize}
\item  modeling of halos,
\item  solar system tests.
\end{itemize}
\item Quantization of the Hamiltonian dynamics for the darkon fluid as a whole or for the cosmological dynamics.
\item Answer the question whether the present model is the nonrelativistic limit of the covariant theory in a strict sense.
\item Discuss higher order derivative terms within the covariant theory.
\end{itemize}

These problems are currently under investigation. We hope to be able to present our results soon.

\eject

\end{document}